\documentclass{elsarticle}
\usepackage{graphicx}
\usepackage{color}
\usepackage{lineno}
\usepackage{upgreek} 
\usepackage{latexsym}  
\usepackage{amssymb}

\begin{document}

\begin{frontmatter}

\title{An evaluation of the exposure in nadir observation of the JEM-EUSO mission}

\author{
J.H.~Adams Jr.$^{md}$,
S.~Ahmad$^{ba}$,
J.-N.~Albert$^{ba}$,
D.~Allard$^{bb}$,
M.~Ambrosio$^{df}$,
L.~Anchordoqui$^{me}$,
A.~Anzalone$^{dh}$,
Y.~Arai$^{eu}$,
C.~Aramo$^{df}$,
K.~Asano$^{es}$
% ...
}
\address{
% ...
$^{ba}$ Laboratoire de l'Acc\'el\'erateur Lin\'eaire, Univ Paris Sud-11, CNES/IN2P3, Orsay, France\\
  $^{bb}$ APC, Univ Paris Diderot, CNRS/IN2P3, CEA/Irfu, Obs de Paris, Sorbonne Paris Cit\'e, France\\
% ...
$^{df}$ Istituto Nazionale di Fisica Nucleare - Sezione di Napoli, Italy\\
% ...
$^{dh}$ INAF - Istituto di Astrofisica Spaziale e Fisica Cosmica di Palermo, Italy\\
% ...
$^{es}$ Interactive Research Center of Science, Tokyo Institute of Technology, Tokyo, Japan\\
% ...
$^{eu}$ High Energy Accelerator Research Organization (KEK), Tsukuba, Japan\\
% ...
$^{md}$ University of Alabama in Huntsville, Huntsville, USA\\
  $^{me}$ University of Wisconsin-Milwaukee, Milwaukee, USA\\
% ...
  \thanks{corr}{Corresponding authors: Mario Bertaina 
    {\it (E-mail address: bertaina@to.infn.it)}, Kenji Shinozaki 
    {\it (E-mail address: kenjikry@riken.jp)}}
}

\begin{abstract}
  We evaluate the exposure during nadir observations with JEM-EUSO, the 
  Extreme Universe Space Observatory, on-board the Japanese Experiment 
  Module of the International Space Station.  Designed as a 
  mission to explore the extreme energy Universe from space, JEM-EUSO 
  will monitor the Earth's nighttime atmosphere to record the ultraviolet 
  light from tracks generated 
  by extensive air showers initiated by ultra-high energy cosmic rays. 
  In the present work, we discuss the particularities of 
    space-based observation and we compute the annual exposure in nadir 
    observation. The results are based on studies of the expected 
    trigger aperture and observational duty cycle, as well as, on the 
    investigations of the effects of clouds and different types of background
    light. We show that the annual exposure is about one order of magnitude 
    higher than those of the presently operating ground-based observatories.
\end{abstract}
\end{frontmatter}


\section{Introduction}
\label{introduction}
The origin and nature of Ultra-High Energy Cosmic
Rays (UHECRs) remains unsolved in contemporary astroparticle 
physics. Possible indications of sources or excesses in the arrival 
direction distribution of UHECRs  have been claimed by ground-based 
experiments, though not fully confirmed \cite{Takeda,Abbasi,augercr}.  
In order to be identified from Earth, extremely powerful 
sources capable of accelerating cosmic rays up to ultra-high energies 
must be within a limited range of distances set by the 
Greisen-Zatseptin-Kuz'min (GZK) effect \cite{GZK,GZK2}.

Since the distribution of matter within the GZK range is 
inhomogeneous and anisotropic,
one would expect UHECR arrival directions to exhibit a 
corresponding anisotropy. To identify the sources of UHECRs, 
measurements of the energy spectrum and arrival directions 
with high statistics are essential. This is 
rather challenging because of the extremely low flux 
of a~few~per~km$^2$ per century at extreme energies such as 
$E > 5 \times 10^{19}$~eV. The observational exposure
is, therefore, a critical factor.

JEM-EUSO (the Extreme Universe Space Observatory on-board the 
Japanese Experiment Module) \cite{Ebisuzaki,Takahashi,Casolino} 
on the International Space Station (ISS) \cite{ISSlimit} is an 
innovative space mission. Looking down the Earth 
from space, it utilizes the atmosphere as a detector of 
cosmic ray air showers with the aim of significantly increasing 
the exposure to UHECRs compared to the largest ground-based air 
shower arrays presently in operation \cite{TA,Yakutsk,Auger}.
The JEM-EUSO telescope will be accommodated on the Exposed 
Facility (EF) on the JEM module 
{\it Kibo} \cite{kibo} of the ISS. The scientific objectives 
include charged particle astronomy and astrophysics, as well as, 
other exploratory objectives \cite{GMT} such as the detection of 
extreme energy gamma rays and neutrinos. The JEM-EUSO 
telescope exploits the fluorescence light that 
is emitted during the development of the Extensive Air Shower (EAS), 
initiated by a primary cosmic ray particle in the atmosphere to 
estimate the particle's energy and arrival direction. This is an 
established technique that has been employed by several 
ground-based UHECR observatories \cite{TA,Auger,FlyEye,HiRes},
but never in space-based observations.

The estimation of the exposure of a space-based experiment such as
JEM-EUSO requires accounting for: a) the characteristics of the EAS 
development in the atmosphere as observed from space, b) the properties of the
telescope, including its orbit and Field of View (FoV), c) the various 
sources of steady background like night-glow and moonlight, 
d) the overall optical transmission properties of the atmosphere, in
particular the possible presence of clouds, and e) the 
effect of anthropogenic light, atmospheric flashes such as Transient Luminous 
Events (TLEs) and meteors. Items a) and b) are the principal factors 
determining the threshold in energy and maximum 
aperture of the telescope. Item c) limits the 
observational duty cycle of the mission. Items d) and e) 
affect the instantaneous aperture of the telescope. 
In the following sections, all of these aspects will be reviewed. 

The outline of this article is as follows:
in Section~\ref{apparatus}, we
summarize the key aspects of the JEM-EUSO mission.
In Sections~\ref{background} and~\ref{sec:clouds}, we discuss the estimation
of the observational duty cycle, local light effects and the role of clouds. 
Section~\ref{trigger}
is devoted to the trigger architecture and EAS
simulation. Computation of the aperture 
for both clear and cloudy conditions will be described in
Section~\ref{sec:aperture}. The exposure in the nadir observation
and its
uniformity will then be derived in Section~\ref{exp-exp}. 
The paper concludes with a discussion of the results 
and a summary in Sections~\ref{discussion} and ~\ref{summary}.


\section{JEM-EUSO telescope and its observation principle}
\label{apparatus}

The JEM-EUSO telescope \cite{Kajino} consists of four 
principal parts: the photon collecting optics \cite{Alex}, the Focal 
Surface (FS) detector \cite{Kawasaki}, the electronics
\cite{Casolino2}, and the mechanical structure \cite{MRicci}.  
The main parameters of JEM-EUSO telescope are summarized in
Table~\ref{tab1}.
The telescope optics consists of three double-sided 
curved circular Fresnel lenses with 2.65~m maximum diameter. 
The minimum diameter of the lenses is 1.9~m owing to cuts on opposite sides.
This shape is referred to as `side-cut' 
and is required to satisfy constraints  
of the H-IIB Transfer Vehicle (HTV)
{\it Kounotori} \cite{kibo} which will transport the JEM-EUSO telescope 
to the ISS. The UV photons are focused onto the FS which consists of 137 
Photo-Detector Modules (PDMs). Each PDM comprises of a 
$3\times 3$ set of Elementary Cells (ECs). Each EC
is formed by a $2\times 2$  array of Multi-Anode PhotoMultiplier Tubes 
(MAPMTs) --
Hamamatsu Photonics K.K. R11265-03-M64 -- with $8\times 8$ (= 64) 
pixels. Each pixel has a spatial resolution of 
$0.074^\circ$.  The FS detector converts 
photons into electrical pulses with $\sim$2~ns width, which are 
counted by the electronics
during a Gate Time Unit (GTU) of 2.5~$\mu$s. 

The imaging part of the telescope is an extremely fast, highly
pixelized, large-aperture, and wide-FoV digital camera. It is sensitive 
to near UltraViolet (UV) wavelength band between 
about 300 and 430~nm with single photon counting capability.
The telescope records the spatial and temporal profile of the UV
light emitted as an EAS develops in the atmosphere. 

\begin{table*}
  \caption{Parameters of the JEM-EUSO telescope. The values 
    in parenthesis apply at the edge of the FoV, 
      otherwise  at the the center of the FoV. The ensquared 
    collection efficiency is the ratio of the number of photons 
    focused within a pixel area to those incident on the entrance 
    aperture of the optics. The ensquared energy is the ratio of 
    photons focused within the area of a pixel to those reaching 
    the FS.}
  \label{tab1}
  \begin{center}
    \begin{tabular}{@{}l|r|l@{}}
      \hline
      Parameter & \multicolumn{1}{|l|}{Value} & Note\\
      \hline
      {\bf Optics} & & \\
      Optical aperture & 4.5 m$^2$ & baseline\\
      Ensquared collection efficiency & 35\% (15\%) & for $\lambda$ = 
      350~nm\\
      Ensquared energy  & 86\% (80\%) & for $\lambda$ = 350~nm  \\
      Optical bandwidth & 300--430~nm & \\
      Field of view & 0.85~sr & \\
      Observational area & $1.4 \times  10^5$~km$^2$ & for 
      $H_0 = 400$~km\\
      \hline
      {\bf FS detector and electronics} & & \\
      Number of pixels & 3.2 $\times $ 10$^5$ \\
      Spatial angular resolution & 0.074$^\circ$ \\
      Pixel size at ground & 0.51~km (0.61~km) & for $H_0 = 400$~km\\
      Quantum efficiency & 41\% & $\lambda=350$~nm\\
      Collection efficiency & 80\% \\
      Cross talk  &     $<2\%$\\
      Transmittance of UV filter & 97\% & for $\lambda$ = 350~nm \\
      Sampling time & 2.5~$\mu$s & \\
      \hline
    \end{tabular}
  \end{center}
\end{table*}

Since the intensity of the observed light
depends on the transmittance of the atmosphere, the cloud 
coverage and the height of the cloud-tops, JEM-EUSO is equipped with an Atmospheric Monitoring (AM) system \cite{Andrii}. 
To characterize the atmospheric conditions as precisely as possible and 
thus determine the effective observation aperture with high accuracy, the AM system consists of an InfraRed (IR) camera and a LIDAR (LIght Detection And Ranging) system. 
Additional information on atmospheric conditions 
is also extracted from the UV data acquired continuously by the
JEM-EUSO telescope itself.

\begin{figure}[!t]
  \begin{center}
   \includegraphics[width=76mm]{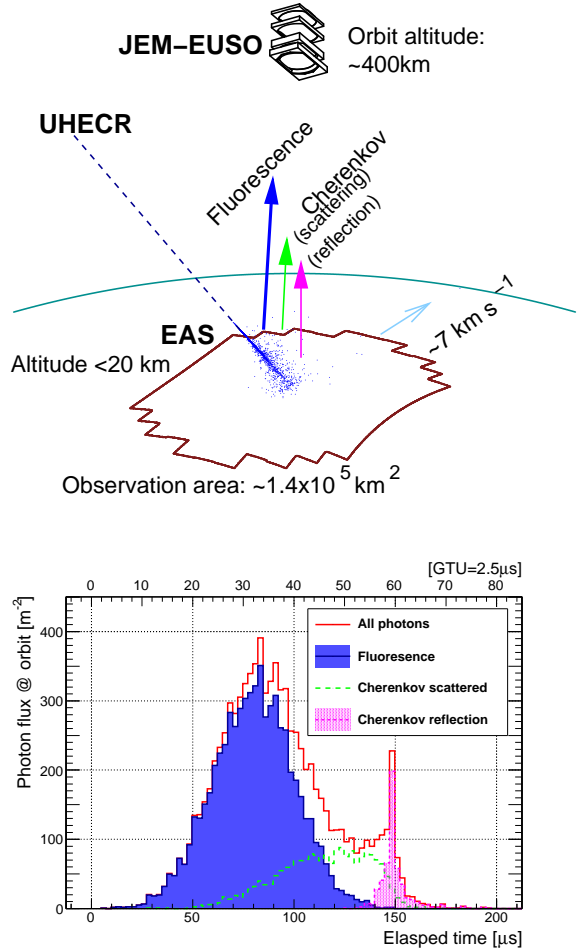}
  \end{center}
  \caption{Top: Illustration of UHECR observation principle in the
    JEM-EUSO mission.  For the telescope at $H_0 \sim 400$~km altitude, 
    the main signals are fluorescence photons along the EAS track and 
    Cherenkov photons diffusely reflected from the Earth's surface. 
    Bottom: Components of the photon signal at the entrance aperture 
    for a standard EAS with $E = 10^{20}$~eV and  $\theta = 60^\circ$ 
    as simulated by ESAF (see Section~\ref{trigger}). 
  }
  \label{art}
\end{figure}
  
The top panel of Figure~\ref{art} illustrates the UHECR observation 
principle in the JEM-EUSO mission. From an orbit at the altitude of 
$H_0 \sim 400$~km, the JEM-EUSO telescope detects fluorescence and 
Cherenkov light from EAS. The fluorescence light is 
emitted isotropically along the EAS track and is 
observed directly. Since the Cherenkov light is forward-beamed, it
is observed either because of scattering in the atmosphere or 
because of diffuse reflection from the surface of the 
Earth or a cloud-top. 
The latter is referred to as `Cherenkov mark' and 
it provides additional information about the shower geometry. 
A $10^{20}$~eV UHECR produces an EAS with $\mathcal{O}(11)$  
particles in the region where the shower reaches its maximum size. 
Secondary charged particles, predominantly electrons, 
excite atmospheric nitrogen molecules that cause UV fluorescence light 
emitted at characteristic 
lines in the band $\lambda \sim 300-430$~nm. 
The fluorescence yield has been 
intensively studied by many groups and is found to be $\sim 3-5$ 
photons~m$^{-1}$~per~electron \cite{flu,phil-flu}. 
During the development of a 10$^{20}$~eV EAS, 
an order of $10^{15}$ photons are 
emitted.
 Seen from $\sim 
400$~km distance, the solid angle subtended by a telescope with a few-m$^{2}$ 
aperture is $\sim 10^{-11}$~sr. 
This implies several thousands of photons reach the entrance aperture of the telescope
under clear atmospheric conditions.

The arrival time distribution of photons at the entrance aperture, is 
presented in the bottom panel of Figure~\ref{art}. The fluorescence 
light is the dominant component, with smaller contributions coming
from reflected and back-scattered Cherenkov light. 
Since fluorescence light dominates the signal, the energy can be determined
with only small corrections for the Cherenkov component.  From 
$H_0 \sim 400$~km, 
the brightest part of the EAS development, which occurs 
below $\sim 20$~km altitude, appears always at an almost constant distance, 
for a fixed location of the EAS in the FoV,
regardless of the direction of the 
EAS, strongly reducing the proximity effects. These are advantageous 
characteristics of space-based experiments. In a sense, 
JEM-EUSO functions as a huge time projection chamber. 
In addition, Cherenkov light reflected from surface of the 
ground or cloud-top is useful for providing a time mark for 
the terminus of the shower. 
 
The orbit of the ISS has an inclination 51.6$^\circ$ and $H_0$ can 
range between 278~km and 460~km according to the operational limits 
\cite{JAXA}. The sub-satellite speed of ISS and the orbital period 
are $\sim 7$~km~s$^{-1}$ and $\sim 90$~min, respectively. Apart 
from effects of orbital decay and operational boost-up, the ISS 
orbit is approximately circular. $H_0$ varies on long-time scale. 
In the present work, we assume $H_0 = 400$~km as a constant value.

The ISS attitude is normally +XV V +ZLV attitude \cite{attitude}
and deviates from it only for
very short periods. +XV V +ZLV is the operational attitude for
JEM-EUSO.
The JEM-EUSO telescope is designed to 
point to nadir, referred to as `nadir mode', as well as to tilt 
astern to the direction of the motion, referred to as `tilt mode'. 
In the following argument, we focus on the case of nadir observation. 

\begin{figure}[!t]
  \vspace{5mm}
  \begin{center}
  \includegraphics[width=78mm]{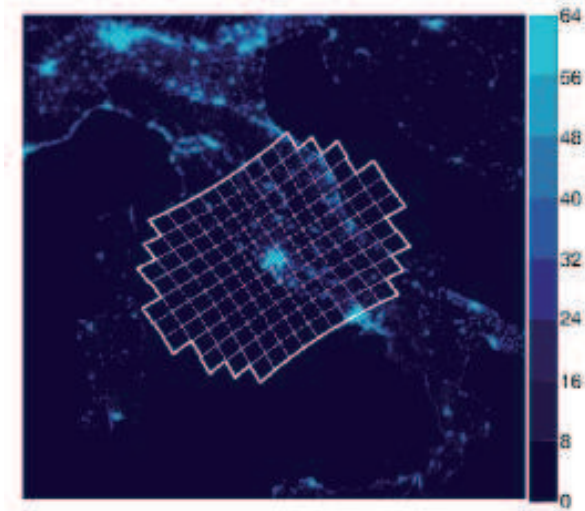}
  \end{center}
  \caption{Observation area of JEM-EUSO telescope flying over central Italy. 
    The background in the map shows visible light distribution obtained
    by DMSP data. The scales denote the values in DMSP units
    (see Section~\ref{background} for details).}
  \label{fig:city-light}
\end{figure}

The observation area of the Earth's surface is essentially determined by 
the projection of the FoV of the optics and the area of the FS. The FoV of the 
optics is estimated using ray tracing simulations \cite{Alex,Kazuhiro}. 
Ray tracing can be used to map the focal surface onto the surface of the Earth
as shown in Figure~\ref{fig:city-light}.

Figure~\ref{fig:city-light} shows the outline of the focal surface mapped onto 
the surface of the Earth (solid 
curves) and the maps of individual PDMs onto the Earth's surface (dashed curves) 
for the case when the 
ISS is located at $H_0 = 400$~km. The background 
in the figure represents
the annual average intensity of light pollution measured by
the DMSP satellite (see the next section for further details).

The dimensions of the FoV are $\sim 64^\circ$ and $\sim 45^\circ$ on the 
major and minor axes, respectively. For these axes, the projected lengths on 
Earth's surface are $\sim 500$~km and $\sim 330$~km, respectively for $H_0 = 400$~km.
The 
effective solid angle $\Omega_{\rm FoV}$ is $\sim 0.85$~sr.
For the planned layout of PDMs on the FS, the size of the observation area
$S_{\rm obs}$ is a function of $H_0$ expressed by:

\begin{eqnarray}
  S_{\rm obs}[{\rm~km}^{2}] & \equiv & \Omega_{\rm FoV} {H_0}^2 \\ \nonumber
  & = & 1.4 \times 10^{5} \cdot 
  {\displaystyle \left(\frac{H_0}{400[{\rm km}]}\right)^2}.
\end{eqnarray}
It is worth noting that the wide FoV allows measurement of the entire
EAS development from the early stage until it fades out or impacts the
Earth. This is especially important for EASs from large
zenith angles and gamma ray or neutrino induced EASs \cite{Daniel}.

\section{Background and observation efficiency} 
  
\label{background}

The UV tracks of an EAS must be discriminated from the UV background.
One parameter essential to estimate the JEM-EUSO exposure is the fraction 
of time during which EAS observation is not hampered by the brightness of 
the atmosphere. We denote the main component of the intensity of diffuse 
background light at the JEM-EUSO telescope, $I_{\rm BG}$, a quantity which
is variable over time. We define the observational duty cycle, 
$\eta$, the fraction of time 
during which the background intensity is lower than 
a given value $I_{\rm BG}^{\rm thr}$. 
We have: 
\begin{equation}
  \eta  \left(<I_{\rm BG}^{\rm thr}\right) 
  = {\displaystyle \eta_{\rm night} 
    \int^{I_{\rm BG}^{\rm thr}}_0 
    p(I_{\rm BG}) d I_{\rm BG}},
\label{eqn:obsduty}
\end{equation}
where $\eta_{\rm night}$ is the nighttime fraction 
and $p(I_{\rm BG})$ 
is the probability density function of $I_{\rm BG}$ over the nighttime
defined as the absence of Sun in the visible sky 
at the orbit level. 
This requires the zenith angle of the Sun to be 
greater than $109^\circ$ for $H_0 = 400$~km and results in 
$\eta_{\rm night} = 34\%.$ 

Different sources are responsible for lighting the atmosphere in the JEM-EUSO 
FoV, including terrestrial sources like
night-glow, TLEs, and local light such as  
city lights, as well as, extraterrestrial light scattered in the atmosphere, such as 
moonlight, zodiacal light, and integrated star light.
While most of these sources
affect the entire FoV,
local light only affect portions of the FoV. Therefore, the
contribution from local light will be considered separately as a term 
that decreases the instantaneous aperture of the apparatus.

Moonlight is the largest background component.
We estimate moonlight contamination from the phase of 
the Moon together with its apparent position as seen from the ISS. 
In our approach, the ISS trajectory provided by NASA SSCweb \cite{NASASSC} 
is traced with 1-min time steps and the moonlight at the top of the 
atmosphere is estimated according to \cite{Montanet}.
For every position of the ISS in the period from 2005 till 2007, 
the zenith angle of the Sun, and that of the Moon,  
$\theta_{\rm M}$, as well as the Moon phase angle, $\beta_{\rm M}$, 
are calculated. Background level from reflected moonlight, 
$I_{\rm M}$, is evaluated using a  modified version  
of the technique 
described in \cite{Montanet}. The UV flux from the full 
Moon is estimated   
according to the magnitudes and color index from \cite{apq}. The apparent 
visible magnitude at 550~nm of the full Moon is $V=-12.74$, while the color 
index $U-V = 1.38$. This yields an ultraviolet magnitude 
of $U=-11.36$ at 360~nm. This corresponds to $2.7 \times 10^5$ photons 
m$^{-2}$~ns$^{-1}$ for $\lambda = 300-400$~nm.
The $\beta_{\rm M}$-dependence of magnitudes is well approximated in 
\cite{Krisciunas}. 

The mean albedo of the Earth evaluated from direct satellite 
measurements has a value close to 0.31 \cite{Frida}. Note that 
such measurements do not distinguish the presence of cloud or other
effects that may increase the background intensity. Taking into 
account the wavelength dependence of the reflectivity \cite{Koelemeijer} as well,  
we set a conservative value of 0.35 for the reflectivity in the range 
$\lambda = 300-400$~nm. Assuming that the radiance of moonlight on the 
top of the atmosphere is diffusely scattered, the overall intensity of 
the backscattered moonlight is estimated to be:  

\begin{equation}
  I_{\rm M}  =  \left[1.6 \times 10^4 \cdot 
    10^{-0.4 \times (1.5 \cdot |\beta_{\rm M}| + 4.3 \times 10^{-2} 
      {\beta_{M}}^4 )} \right]  \cos \theta_{\rm M}, 
\label{eqn:moonshine}
\end{equation}
where $\beta_{\rm M}$ is in radians and $I_{\rm M}$ has units of photons
m$^{-2}$~ns$^{-1}$~sr$^{-1}$. The overall background intensity, 
$I_{\rm BG}$, is given by:

\begin{equation}
I_{\rm BG} = I_{\rm M}(\theta_{\rm M},\beta_{\rm M}) + I_0 ,
\end{equation}
where $I_0$ represents the stable contribution of UV background
created mainly by night-glow. 
In this calculation, $I_0$ is assumed to have a constant value of 
500 photons m$^{-2}$~sr$^{-1}$~ns$^{-1}$ in the range 
$\lambda = 300 - 400 $~nm 
\cite{Garipov,NIGHTGLOW,BABY}. Taking into account the responses of the optics
and FS detectors, the average background level on the MAPMTs corresponds to 
$\sim$ 1.1 photoelectrons GTU$^{-1}$~per~pixel.

\begin{figure}[!t]
  \vspace{5mm}
  \begin{center}
  \includegraphics[width=76mm]{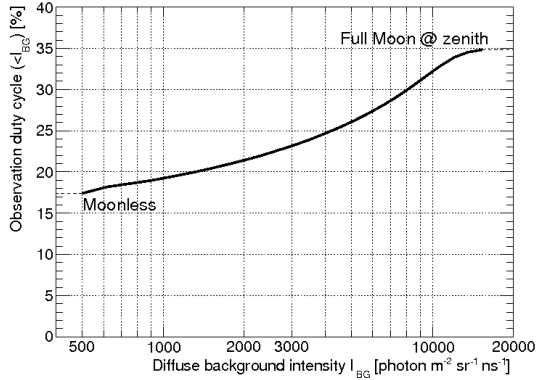}
  \end{center}
  \vspace{-0.5cm}
  \caption{Observational duty cycle $\eta$ as a function of threshold
    background level $I_{\rm BG}^{\rm thr}$. 
}
  \label{fig:duty}
\end{figure}

Figure~\ref{fig:duty} shows the observational duty cycle as a function of
the accepted background level according to Equation~(\ref{eqn:obsduty}).
The fraction of time during which $I_{\rm BG}$ is less than 
1500 photons m$^{-2}$~sr$^{-1}$~ns$^{-1}$ is 20\%--21\%
resulting in an average background level of 
550 photons m$^{-2}$~sr$^{-1}$~ns$^{-1}$ under this condition.
The threshold of 1500 photons m$^2$ ns$^{-1}$ sr$^{-1}$ 
is chosen as a reference. This corresponds to the decrease of signal to noise 
ratio by a factor of $\sim \sqrt{3}$ in comparison to the case of $I_{\rm BG}$ 
= 500 m$^2$ ns$^{-1}$ sr$^{-1}$, which still guarantees EAS observation. 
The effects of variable background level will be discussed in 
Section~\ref{exp-exp}. Of course, the operation of the instrument is 
not limited by this value, so this is a conservative estimate for the 
highest energies, where measurement can be performed even in a higher 
background condition. In the following discussion, we assume 
$\eta_0\sim$20\% as the reference value for the observational duty cycle.

In addition to the diffuse sources of background, there are intermittent local 
sources such as lightening flashes, auroras, or city lights.
In the cases of lightening and TLEs, estimates of the reduction in 
observational duty cycle and instantaneous aperture are performed  
assuming the rate of events detected by Tatiana satellite 
\cite{Garipov}. 
    We further assume that the EAS measurement is not possible in all 
    PDMs as long as the location of the event is within the JEM-EUSO FoV. 
    For a conservative estimation, we applied $\sim 70$ s 
to all events, 
    that corresponds to the maximum time a light source takes to traverse 
    the major axis of the FoV.
Even with these 
extremely conservative assumptions, the overall effect is less than 
$\sim 2\%$. Moreover, as lightening is very often associated with 
high clouds, most of this effect is already included in cloud 
inefficiency, as described in the next section.

To estimate the reduction in observational duty cycle and aperture due
to the occurrence of auroras, we used the $K_{\rm p}$ index to describe the
geomagnetic activity, as well as, the geomagnetic latitude and longitude of
ISS during years 2001 and 2006. These two years were selected as they were
close to solar minimum and solar maximum, respectively. In the
estimation, it was assumed that no measurement can be performed when the
$K_p$ index for ISS geomagnetic latitude is equal or higher than Auroral
Boundary Index \cite{aurora}. Even in the case of maximum solar activity the
effect is of the level of $\sim 1\%$.  

To evaluate the effect of the stationary light sources on the Earth,
which are mainly anthropogenic, we use the Defense Meteorological 
Satellite Program (DMSP) \cite{NASANOAA} database.
Annual averages of light intensities for cloud-free
moonless night are used to      
estimate the presence of local light along
the ISS trajectory. The DMSP data provide the light 
intensity in 64 different levels on a 30-arcsecond grid in latitude 
and longitude in the wavelength range 350~nm--2~$\mu$m. 
The units are arbitrary, with equally spaced steps.
The stationary background is dominated by visible light.  
As an example, the average level of background 
around central Italy is shown in Figure~\ref{fig:city-light}.
As the trigger system in JEM-EUSO works at the PDM level (see
Section \ref{trigger}), we discuss the impact of local light
at PDM level here. 

Table~\ref{DMSPtab} summarizes the results on the visible intensity from DMSP data over the region 
between 51.6$^\circ$S and 51.6$^\circ$N latitudes which is covered by ISS 
trajectory.
\begin{table}
  \label{tab:tat}
  \caption{Fraction of pixels with visible light intensities $\le$3, $>$3 and $>$7 (DMSP units)
    in spatial resolutions of 
    DMSP (left columns) and of JEM-EUSO PDM (right columns).}
  \begin{center}
  \begin{tabular}{c|ccc|ccc}
    \hline
       & \multicolumn{3}{c|}{DMSP pixel resolution} & 
    \multicolumn{3}{c}{JEM-EUSO PDM resolution}\\
    \cline{2-7}
    & $\le 3$ &  $> 3$ & $> 7$ & $\le 3$ & $> 3$ & $> 7$   \\
    \hline
    All   & 96\% &   4\% &   2\% & 87\% &  13\% &   7\%  \\
    Land  & 85\% &  15\% &   6\% & 58\% &  42\% &  25\% \\
    Ocean & 99.8\% & 0.2\% & 0.1\% & 99.1\% & 0.9\% & 0.4\%\\
    \hline
  \end{tabular}
  \end{center}
  \label{DMSPtab}
\end{table}
From the DMSP data, which have their own spatial resolution, the average 
intensity is 2.6 in DMSP units.
This value is mainly determined by the background over the ocean which 
represents
72\% of the JEM-EUSO observational region. In the following, we make the 
conservative assumption that no measurement of EASs 
is performed if, in a region viewed by a PDM,
there is at least one pixel which detects a light intensity which exceeds 
the average level by a factor of 3 or more 
(higher than 7 in a DMSP units). With this assumption, the inefficiency of the 
instantaneous aperture is of the order of $\sim$7\%. It is important to 
remember that Tatiana measurements \cite{Garipov} -- without focusing 
optics -- indicate a 2--3 times higher intensity in UV above big cities such 
as Mexico City and Houston compared to the average background level over the 
ocean.
Finally, by combining the above estimations for lightnings ($\sim$2\%),
auroras ($\sim$1\%) and DMSP data the overall loss of coverage is 
$f_{\rm loc}$ $\sim$10\%.

\section{Climatological distribution of clouds}
\label{sec:clouds} 

In the case of space-based observation, reconstructing an EAS event
is feasible, 
even in the presence of clouds, if the EAS maximum is sufficiently
above the cloud-top altitude,
$H_{\rm C}$ \cite{Takahashi,AbuZayyad}. In some cases, the presence of 
specific cloud types may even be an advantage 
(see Section~\ref{sec:aperture}), which is contrary to 
ground-based observation.
An optically thick cloud represents a very uniform layer which enhances the 
intensity of the 
Cherenkov mark and gives a brighter end point of the track. Of course, 
the cloud-top height should be known with reasonable uncertainty 
($\sim 0.5-1$~km), and for that, the AM system is used. In the case of 
optically thin clouds or very inclined showers, however, 
the Cherenkov mark is not well defined. 
Therefore, it is mandatory to develop alternative 
reconstruction algorithms which do not rely on the detection of the 
Cherenkov mark \cite{Fenu}. 
The combined use of algorithms based on different approaches
on an event-by-event basis helps to prevent, 
or at least tag, misreconstructed events.
Thin clouds with optical depths $\tau_{\rm C} <$1  (typically cirrus) 
may affect the estimation of the energy, but the arrival direction 
can be determined with acceptable uncertainty. 
In such a case, the estimated energy is likely to be lower than 
the true one, adding to a given reconstructed energy bin an event whose 
true energy is in fact larger, and thus whose angular deflection is a 
priori smaller -- not the opposite. Even though such a situation may alter 
the quantitative estimates of the anisotropy as a function of energy, 
some anisotropy analyses will still be interesting to perform with such 
events, notably those assessing a lower limit on anisotropy.
Optically thick clouds, with $\tau_{\rm C} >$1, strongly 
influence the measurement only if they are located at high altitudes.  
For example, EASs from a 60$^\circ$ zenith angle and energy
$\sim 10^{20}$~eV reach their maxima at an altitude around $\sim 6.5$~km, 
much higher than the typical range  of stratus. The effect of clouds is, 
therefore, to limit the instantaneous aperture by obscuring portions of 
the FoV.

In order to quantify the effect of cloud contamination, a study of the 
climatological distribution of clouds, 
as a function of cloud-top altitude, optical depth,  and geographical 
location 
has  been performed using the meteorological databases TOVS, 
ISCCP and CACOLO.

The NASA project TOVS (TIROS Operational Vertical Sounder) \cite{TOVS}
on board NOAA's TIROS series of polar orbiting satellites provides data 
with a good
spectral distribution, as well as optical depth and altitude of 
clouds, which are obtained applying a radiative transport model \cite{TOVS}.
In this study, data from
1988 to 1994 have been used, including both land and ocean data.

The International Satellite Cloud Climatology Project (ISCCP)
\cite{ISCCP} was established in 1982
to collect and analyze satellite radiance measurements needed to infer 
the global
distribution of clouds, their properties and their diurnal,
seasonal and inter-annual variations. The ISCCP has developed cloud 
detection
schemes using visible and IR window radiance (IR during nighttime and
daytime, and visible during daytime). The data from 1983 to 2008 have
been used in this analysis. Data are given on a 2.5-degree grid in
latitude and longitude.

The CACOLO (Climatic Atlas of Clouds Over Land and Ocean data) database
\cite{CACOLO}
presents maps introduced in the atlases of cloud climatological data
obtained from visual observations from Earth. Most data are given at
a 5-degree resolution in latitude and longitude. The land data are
based on analysis of visual cloud observations performed at
weather stations on continents and islands over a 26-year period
(1971--1996). The ocean maps are
based on analysis of cloud observations made from ships over a
44-year period (1954--1997).

Systematic differences between these databases have been evaluated. As 
previously explained, the ISCCP and CACOLO data divide the
clouds only in low ($H_{\rm C}<$3.2~km), middle ($H_{\rm C}= 3.2-6.5$~km) 
and high types ($H_{\rm C}>$6.5~km) without distinguishing
according to their optical depths. Some care has to be taken with 
CACOLO results, as these data are based on observations from the ground, 
so the cloud altitude refers to the cloud bottom. 
Since the CACOLO data characterize well the cloud occurrence in the lower 
part of the atmosphere, they nicely complement the observations from space. 

\begin{figure}[!t]
  \vspace{5mm}
  \begin{center}
  \includegraphics[width=76mm]{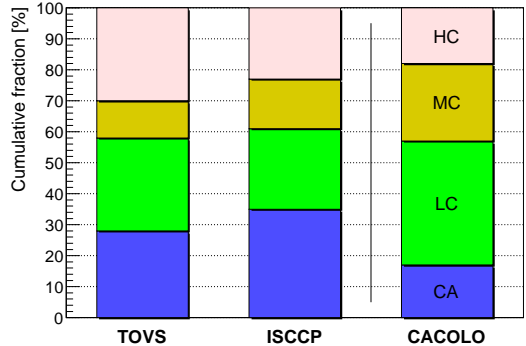}
  \end{center}
  \vspace{-0.5cm}
  \caption{Comparison of the TOVS, ISCCP and CACOLO databases 
    for the relative cloud occurrence in different meteorological 
    situations. The data correspond to daytime, with a weighted average 
    between land and ocean. The classification of the TOVS and ISCCP data 
    is based on the cloud-top altitude while for CACOLO it is based on the 
    cloud-bottom. High clouds are defined by $H_C$ $>$ 6.5 km; Middle 
    clouds by $H_C$ = 3.2 -- 6.5 km; Low clouds by $H_C$ $<$ 3.2 km. The abbreviation are defined as 
    HC: High Clouds; MC: Middle Clouds, LC: Low Clouds and
    CA: Clear Atmosphere.}
  \label{tab4}
\end{figure}

Figure~\ref{tab4} shows a comparison of cloud distribution in the
troposphere among the three datasets 
where land and ocean data are combined in a weighted average.
Only daytime data are shown, since CACOLO employs visual observation.

Despite the fact that significant differences exist
among the three databases for each of the 4 categories shown, 
if one considers the case of good JEM-EUSO conditions, namely low clouds or
clear atmosphere, the three datasets are in reasonable agreement,
with a minimum of 57\% for 
CACOLO to a maximum of 61\% in case of ISCCP. 
The difference among high cloud measurements might be due to the fact that 
CACOLO data are taken by ship and weather stations only in the visual band. 
This could result in a smaller fraction of high clouds, especially in 
presence of low and middle altitude clouds. In contrast, TOVS data are 
taken by satellites, therefore, low and middle altitude clouds 
may be underestimated due to the obscuration by  high clouds. ISCCP data 
show a more uniform cloud occurrence among the different atmospheric
levels most probably because they are taken from satellite in the visual 
and IR bands that enables distinguishing the various levels. 
Since the TOVS data show the highest fraction of high 
clouds, which are the most critical in case of EAS observation from
space, estimates of the fraction of EASs measurable by JEM-EUSO 
using Table~\ref{tab2} can be considered as conservative.

Table~\ref{tab2} reports TOVS data on the occurrence 
of each cloud category during nighttime on the globe and above 
the ocean only. The results apply only to the region 
of the ISS trajectory
and account for the residence time of the ISS as a function of latitude. 

\begin{table}
  \begin{center}
    \caption{Relative occurrence of clouds
        over the ISS orbit, taken from the TOVS database for nighttime,
        are presented as a matrix of cloud-top altitude vs optical
        depth for all location and only ocean.}
      \label{tab2}
      \begin{tabular}{l|cccc}
        \hline
        Cloud-top&\multicolumn{4}{c}{Optical depth $\tau_{\rm C}$}\\ \cline{2-5}
        altitude $H_{\rm C}$ & $<$0.1 & 0.1--1 & 1--2 & $>2$\\ \hline
        &\multicolumn{4}{c}{All data}\\
        $>10$~km    & 1.2\%    & 5.0\% & 2.5\% & 5.0\% \\
        6.5--10~km  & $<0.1\%$ & 3.2\% & 4.2\% & 8.5\% \\
        3.2--6.5~km & $<0.1\%$ & 2.0\% & 3.0\% & 6.0\% \\
        $<3.2$~km   & 31\%     & 6.4\% & 6.0\% & 16\%  \\
        \hline
        &\multicolumn{4}{c}{Ocean data}\\
        $>10$~km    & 0.1\% & 5.0\% & 2.4\% & 4.7\% \\
        6.5--10~km  & 0.1\% & 3.2\% & 4.3\% & 9.2\% \\
        3.2--6.5~km & 0.1\% & 2.1\% & 3.1\% & 5.7\% \\
        $<3.2$~km   & 29\%  & 6.6\% & 6.5\% & 17\%  \\
        \hline
      \end{tabular}
  \end{center}
\end{table}


A comparison between day and night cloud coverage has been 
performed for clouds above land as higher variations are expected in 
comparison with the day-night variation above the ocean. Slight 
differences among tables exist (typically $\sim5\%$), though the 
general trend seems to be independent of the geographical and temporal 
conditions \cite{Garino}. In any case, only nighttime
conditions are relevant for JEM-EUSO. By comparing TOVS results in 
Figure~\ref{tab4} 
and Table~\ref{tab2}, no significant difference seems to exist between the 
high (30\% also in Table~\ref{tab2}) and middle (11\% in 
Table~\ref{tab2}) cloud occurrence in daytime and nighttime. The clear 
atmosphere seems to be slightly more frequent in nighttime compared to daytime.

\begin{figure}[!t]
  \vspace{5mm}
  \begin{center}
  \includegraphics[width=76mm]{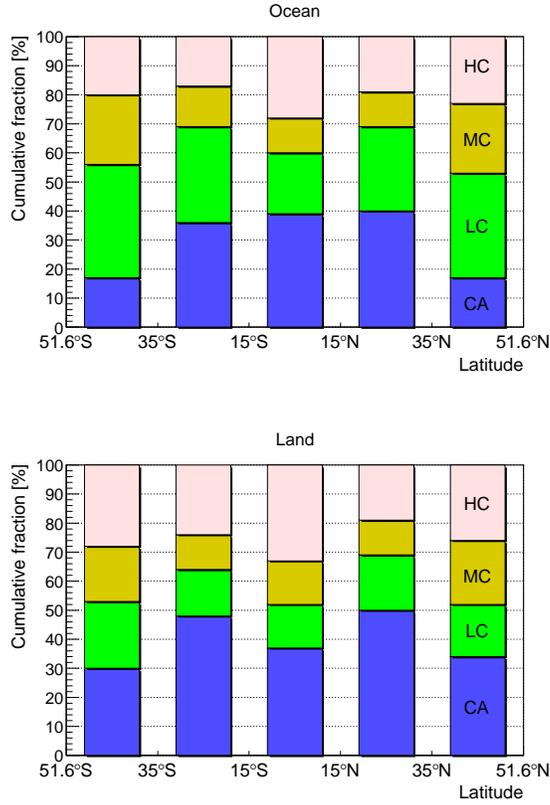}
  \end{center}
  \vspace{-0.5cm}
  \caption{Distributions of the cloud properties over 
    the ocean (top panel) and the land 
    (bottom) are given for 5 latitudinal zones, using 
    ISCCP data. The abbreviation are defined as 
    HC: High Clouds; MC: Middle Clouds, LC: Low Clouds and 
    CA: Clear Atmosphere.
    Data refer to daytime.}
  \label{tab5}
\end{figure}

Finally, the ISCCP data have been used to check the dependence of the above 
results on latitude. Results are summarized in 
Figure~\ref{tab5}.       

In general the occurrence of low clouds 
and clear atmosphere is slightly 
higher over the oceans. It is important to remember that ocean
accounts for 72\% of the time for the ISS orbit. High clouds are 
particularly frequent in the equatorial region.
This is expected, due to the presence of the cloudiness associated to the
Inter Tropical Convergence Zone.

\section{Trigger scheme and shower simulation}
\label{trigger}

Another key parameter in determining the exposure is the 
EAS trigger efficiency of the instrument. This is strongly dependent 
on the characteristics of the optics and FS detector,
as well as the darkness of the atmosphere.
Therefore, the trigger logic explained in the following 
was defined as a trade-off between the available power and telemetry 
budgets of the instrument, the response of the detector, as explained 
later in this section, and the necessity of coping with background 
fluctuations whose excess at tenths microsecond level on an 
MAPMT basis 
could mimic the presence of a signal from an EAS.

To reject the background, the JEM-EUSO electronics employ two 
trigger levels. The trigger scheme relies on the partitioning of the 
FS onto PDMs, which are large enough to contain a substantial part of the 
imaged trace under investigation, as explained below. 

The $1^{\rm st}$ trigger level rejects most of the 
background fluctuations by requiring a locally persistent signal above 
average background lasting a few GTUs. In this trigger level,
referred to as Persistent Track Trigger (PPT), 
pixels are grouped in $3\times 3$ boxes. 
A trigger is issued if for a certain number of consecutive 
GTUs, $N_{\rm pst}$,
there is at least one pixel in the box with an activity 
equal to or higher than a 
preset threshold, $n_{\rm thr}^{\rm pix}$, and the total number of detected 
photoelectrons in the box is higher than a preset value 
$n_{\rm thr}^{\rm box}$.  $N_{\rm pst}$ is set to 5 GTUs in the current 
simulations, while $n_{\rm thr}^{\rm pix}$ and $n_{\rm thr}^{\rm box}$ are set as 
a function of $I_{\rm BG}$ in order to keep the rate of triggers on 
fake events at few Hz per PDM.
For an average background level of 1.1 photoelectron
GTU$^{-1}$ per pixel, $n_{\rm thr}^{\rm pix}$ is set to  
2 and $n_{\rm thr}^{\rm box}$ to 32.
 
The 2$^{\rm nd}$ trigger level \cite{Joerg}, referred to
as the Linear Track Trigger on Cluster Control Board (CCB\_LTT), follows 
the movement of the EAS spot inside the PDM over a predefined time 
window to distinguish the unique pattern of an EAS from the 
background. Starting from the location where the PPT trigger is 
issued, the CCB\_LTT trigger algorithm defines a box of 
$3\times 3$ pixels 
around this trigger seed, then moves the box every GTU.
The box is moved along pre-defined lines to search for 
the direction of the EAS.
The photon count along each line is integrated each 
GTU by summing up the counts $n^{\rm pix}$ of the pixels in the box 
that in such GTU satisfy the condition 
$n^{\rm pix} \ge n_{\rm thr}^{\rm pix}$. 
The integration is performed for 15 consecutive GTUs. If the 
integration along a direction exceeds a prefixed threshold, 
eg. 97 counts under this background level, the 
CCB\_LTT trigger is issued. In order to follow the movement of the 
spot on the focal surface, the speed and the direction in terms of 
detector pixels is calculated according to:
\begin{eqnarray}
  \hat{\theta} &=& 2\arctan\left(\frac{\Delta L}{c\cdot\Delta t}\cdot\sqrt{\Delta x^2+\Delta y^2}\right)\\
  \hat{\varphi} &=& \arctan\left(\frac{\Delta y}{\Delta x}\right),
\end{eqnarray}
where $\hat{\theta}$ and $\hat{\varphi}$ are, respectively, the polar 
angle and azimuthal angle in a spherical coordinate system whose polar axis is 
aligned along
the line of sight of the pixel, 
$c$ is the speed of light, $\Delta x$ and $\Delta y$ are the 
number of pixels crossed in a time $\Delta t$,
and $\Delta L$ is the projected length of the pixel FoV on the 
Earth's surface, which is given in Table~\ref{tab1}.

Since the incoming direction of the EAS is unknown, 
the CCB\_LTT trigger tries 
directions which fully cover the phase space ($\hat{\theta} 
= 5^{\circ}, 10^{\circ},\ldots, 85^{\circ}$ and $\hat{\varphi} = 5^{\circ}, 
10^{\circ},\ldots,355^{\circ}$), which means that the
directions in which the box should move are defined before 
starting the integration for 15 GTUs. 
The integrated count value will have a maximum when the nearest 
direction to the correct one is selected because in this case the 
integrating box will most closely follow the EAS track. The 
$I_{\rm BG}$-dependent threshold on the total number
of counts inside the track is tuned to reduce the fake events to 
a rate of 0.1 Hz on the entire FS. The two trigger levels combined 
operate a reduction in rate by $\sim 2 \times 10^{-7}$ at 
PDM level.
When a trigger is issued, a sufficiently large part of the FoV
(a few PDMs) is acquired in order to image the region around
the EAS track.

The trigger rate for real EASs is less than $\sim1$\% of 
the total trigger rate, depending on the background intensity.
A more comprehensive review of the trigger scheme is given in 
\cite{Catalano}. Besides the fact that the 
threshold in energy is affected by the darkness of the atmosphere 
and the photon collecting power of the telescope, it is important 
to underline here that the trigger system on a space-based 
detector has to be much more selective than a ground-based 
experiment because of telemetry constraints. This limits 
the threshold in energy. On the other side, the fraction of 
scientific data in the sample will be of high quality. Thus, it is 
expected that further quality cuts applied in the 
offline analysis will 
not cause a significant reduction of data.

In order to evaluate the detector response to the EAS 
observation,
we use the Euso Simulation and Analysis Framework (ESAF).
A detailed description of the software can be found in \cite{ESAFpaper}. 
In the following analysis, the 
Greisen-Ilina-Linsley (GIL) function \cite{Ilina} is used 
as parametric generator to
reproduce the profile as a function of slant depth.
The GIL function is optimized to reproduce EAS from hadronic
particles simulated by CORSIKA \cite{Corsika} with the 
QGSJET01 hadronic interaction
model \cite{qgs}. 
Proton showers have been simulated for the analyses presented in this paper. 
This is motivated by the fact that they develop deeper in the 
atmosphere, which results in a higher atmospheric absorption and higher 
cloud impact
\footnote{The EAS observation from space has a better visibility of the 
early stages of the shower development compared to ground-based 
observation. Iron showers tend to cascade
higher in atmosphere compared to proton ones and the apparent length of the 
EAS before impacting on the Earth's surface or on a cloud top
is a bit longer.
Simulation results indicate in case of iron showers that a slightly higher
number of photons reaches JEM-EUSO in comparison to proton showers
with same energy and geometry.
This results in a slightly improved trigger efficiency and increased overall 
exposure in case of iron showers.}.
Therefore, the results that are discussed in
the following sections constitute a conservative estimation on the 
performance of the instrument.

In the present work, the fluorescence yield, which constitutes one of 
the larger uncertainties in energy determination, is taken from 
\cite{ngn}.
In the atmosphere, UV photon propagation is strongly affected by
Rayleigh scattering and absorption by ozone for 
wavelengths $\lesssim 320$~nm. These processes along with the 
atmospheric profile are 
modeled with the LOWTRAN package \cite{lwtrn}. 
  

    The detector simulation includes optical 
    ray tracing, PDM layout on the FS, UV filter, MAPMT performance and 
    trigger algorithm. For the optics response, the simulation code 
    described in \cite{Alex} has been adopted in the present analysis. 
    A parametrization of the MAPMT is included in the electronics 
    simulation. All the effects like quantum efficiency, including the 
    dependence on photon inclination, collection efficiency and cross talk 
    are also taken into account pixel by pixel within one MAPMT as 
    summarized in Table~\ref{tab1}. 

\begin{figure}[t!]
  \begin{center}
    \includegraphics[width=76mm]{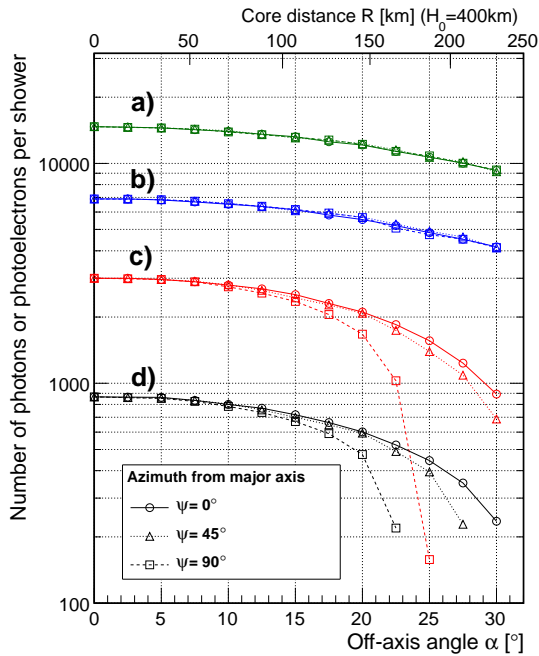}
    \caption{Number of photons and photoelectrons as a function of off-axis 
      angle $\alpha$ of the core location  from nadir, obtained by simulating 100 EASs with $10^{20}$~eV 
      and $\theta = 60^\circ$.  The off-axis angle is the angle between the
core location and the nadir axis.
To demonstrate the azimuthal dependence
      of the optics, three different directions are shown by
      circles $(\psi = 0^\circ)$, triangles  $(\psi = 45^\circ)$
      and squares  $(\psi = 90^\circ)$ where $\psi$ is the angle from
      the major axis of the optics.  The scale on top is the radial distance 
from the center of FoV on the Earth's surface. In the figure, the 
      different stages are compared:   
      a) photons from the shower axis directed toward the JEM-EUSO entrance 
aperture;
      b) photons reaching JEM-EUSO entrance aperture; c) photons reaching the FS 
and d) detected signal (photoelectrons). }
    \label{fig:evol}
    \end{center}
  \end{figure}
  
  Figure~\ref{fig:evol} shows the number of photons and signals
  for the average of 100 EASs with $E = 10^{20}$~eV and $\theta = 60^\circ$.
  To demonstrate the radial and azimuthal dependent response of the detector, 
  mostly
  due to optical vignetting, we simulate EAS with different core locations. 
  The horizontal axis shows the off-axis 
  angle $\alpha$ of the core location with respect to the optical axis 
  corresponding 
  to the direction of nadir. The scale on the top indicates the core 
  distance $R$ from the center of FOV $(\sim H_0 \tan\alpha)$. 
  The detector design is symmetric in each quadrant. The azimuthal 
  dependence on the optics is, therefore, only tested for the case 
  $\psi = 0^\circ$, $\psi = 45^\circ$ and $\psi = 90^\circ$, 
  where $\psi$ is the azimuthal angle of the focusing position
  from the major axis of the FS.

  Independent of $\psi$, the numbers of photons at the entrance aperture, 
  namely 
  stages a) and b) in Figure~\ref{fig:evol}, depend only on the distance 
  between the telescope 
  and EAS, and on an entrance aperture of a given solid angle, so they 
  are roughly 
  proportional to $\cos\alpha$.
We recall here, as explained in Section~\ref{apparatus}, 
that the proximity effects are negligible for a space-based observation.
  The ratio of b) to a) corresponds to the average transmittance for photons 
  reaching JEM-EUSO from the position where
  they are produced either by emission or scattering.  
  As the Cherenkov light has continuous spectrum, we simulated the 
  wavelength range up to 485~nm where the photon detection efficiency
  is negligible.
  Over the FoV of JEM-EUSO telescope ($\alpha \lesssim 30^\circ$,
  or half of FoV for the major axis), 
  the variation of the number of photons at the entrance aperture is within 
  $\sim 1.7$. 

  The decrease from b) to c) indicates effects in the optics, such as  the 
 absorption and scattering of photons in lenses, characteristic aberration 
and obscuration   by the support structure.

  The ratio of d) to c) reflects the efficiency of the FS detector and
  is generally determined by the detection efficiency
  (product of collection efficiency and quantum efficiency) of
  MAPMTs and transmittance of UV filter (Schott BG3 filter
\cite{Schott}). Photons may be lost
  in part when they are focused on void areas such as gaps among PDMs.
  For $\psi = 0^\circ$ the difference between the center and 
  edge of the FoV is a factor of $\sim 2$.  
  A significant dependence on $\psi$ emerges at $\alpha \sim 15^\circ$.
  This is because some of the photons arriving from
  angles close to the minor axis ($\psi = 90^\circ$) are bent to the
  internal lenses on segments that have been removed in side-cut optics. 
  For $\alpha \gtrsim 23^\circ$, no PDM is present on the FS along the minor 
axis of the optics. Note, however,
  that this effect only appears between $\psi \sim  45^\circ$ and 
  $\psi \sim 90^\circ$ and the range of
  corresponding angles in the quadrants where the segment 
  crosses a circular part of the lens.  


 The signal is then amplified using a parametrization of the measured gain 
and the resulting 
 output current is collected and treated by the Front End Electronics. 
 A threshold is set on the MAPMT output current in order to accept or 
 reject the signal count.

The trigger architecture and the parameters used in PTT and 
CCB\_LTT trigger algorithms have been
 optimized using ESAF and stand alone Monte Carlo simulations
 to reduce the fake trigger rate from background
fluctuations to an acceptable level, exploiting the detector 
response such as ensquared energy of the optics,
and detection efficiency, cross-talk among pixels, etc. 

 \begin{figure}
   \vspace{5mm}
   \begin{center}
   \includegraphics[width=76mm]{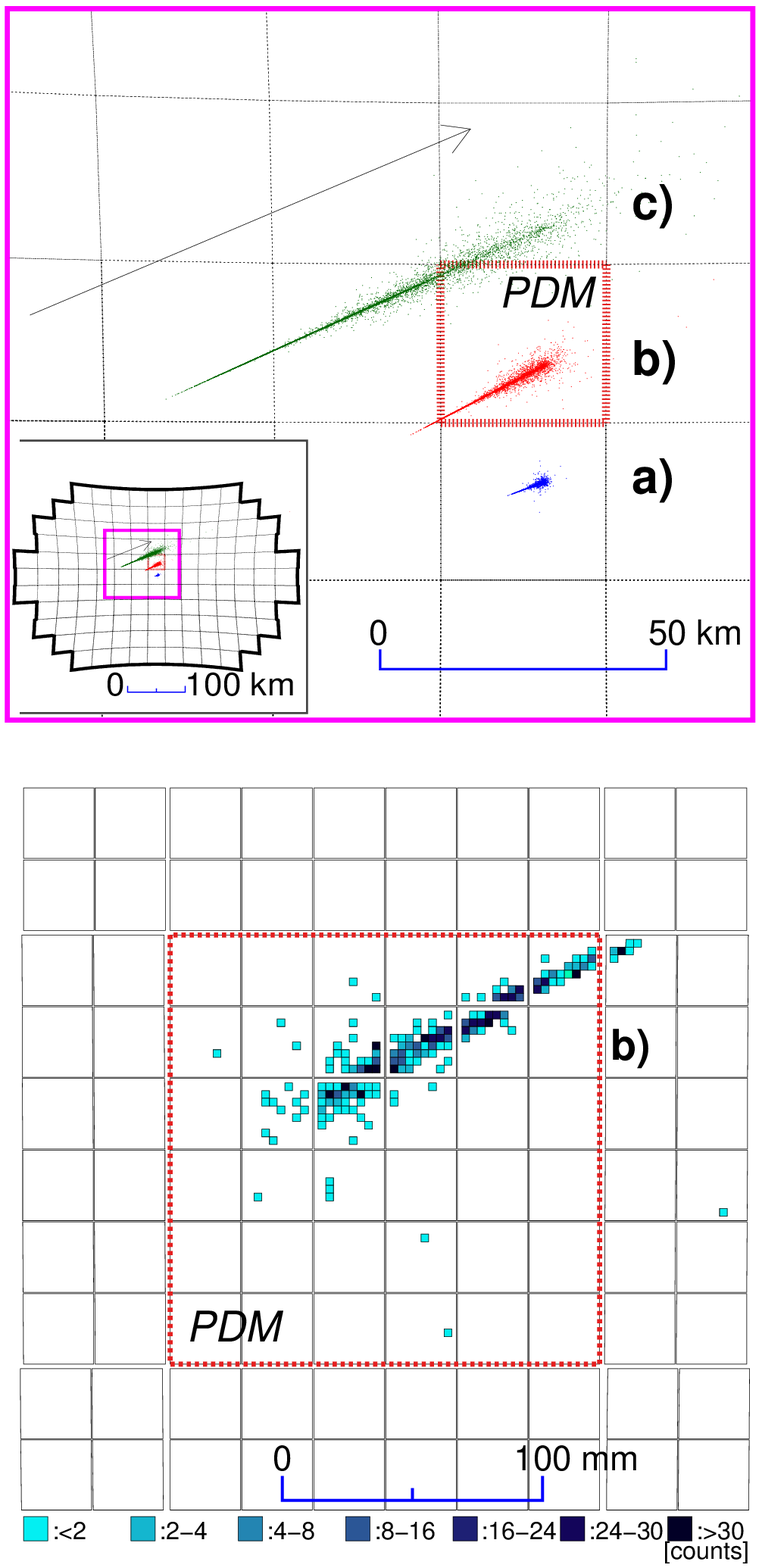}
   \end{center}
   \caption{Top panel shows the projected tracks on the 
Earth's surface for EASs with $E = 10^{20}$~eV 
     and zenith angles of a) $\theta = 30^\circ$, b)
     $\theta = 60^\circ$ and c) $\theta = 75^\circ$.  
The dashed curves indicate the corresponding areas for the 
FoV of individual PDMs.
     In the sub-panel, the corresponding area of the plot is 
represented 
     by solid lines within the entire FoV. 
     Bottom panel shows the image on the FS for the case b).
     The large squares denotes PMTs. The matrix of pixels are 
indicated 
     with the integrate counts in discrete scale. 
The regions enclosed by thick dashed lines in both 
     panels refer to the same PDM.    }
   \label{fig:fsimage}
 \end{figure}

 The ESAF code reconstructs the EAS energy, arrival direction, and 
 longitudinal development of simulated  events, and it is used to check 
 the accuracy of reconstruction. This also provides feedback that is 
 useful for the development of analytical algorithms and hardware 
 to improve the performance of the detector.

 \begin{figure}
   \vspace{5mm}
   \begin{center}
   \includegraphics[width=76mm]{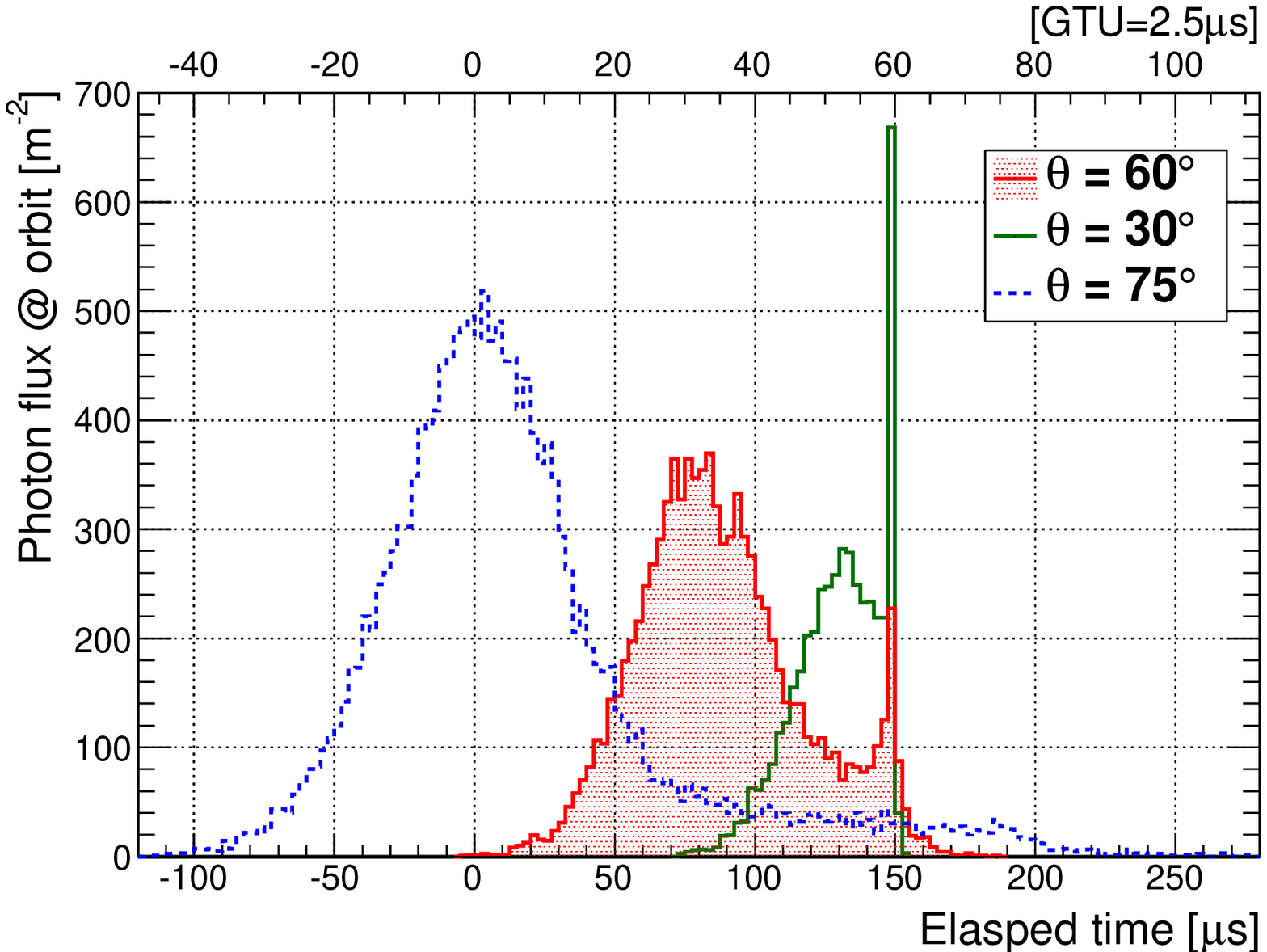}
   \end{center}
   \caption{Arrival time distribution of photons at
      the telescope entrance aperture from the same EASs shown in 
      Figure~\ref{fig:fsimage}. Shaded histogram denotes the
   case of $\theta = 60^\circ$ and those with solid and dashed
   lines are for the cases of $\theta = 75^\circ$ and $\theta 
   = 30^\circ$,  respectively.}
   \label{fig:light-profile}
   \vspace{-0.5cm}
 \end{figure}

  In Figure~\ref{fig:fsimage}, top panel shows the 
  projected tracks on the 
  Earth's surface for EASs with $E = 10^{20}$~eV 
  and zenith angles of a) $\theta = 30^\circ$, b)
  $\theta = 60^\circ$ and c) $\theta = 75^\circ$ along with 
  the map for the entire FoV in the sub-panel.
  Bottom panel shows the image on the FS for the case b)
  in which the integrated counts for each pixel are indicated. 
  The regions enclosed by thick dashed lines in both 
  panels refer to the same PDM. 

 Figure~\ref{fig:light-profile} shows the arrival time distribution of 
 photons at the telescope entrance aperture from the EASs shown in Figure 
 ~\ref{fig:fsimage}. The shaded histogram is for $\theta = 60^\circ$ and 
 those with solid and dashed lines are for $\theta = 75^\circ$ 
 and $\theta  = 30^\circ$, respectively.
 
 Up to zenith angles $\theta \sim$60$^\circ$, 
 the EAS is fully contained in an FoV equivalent to that of one PDM.
 It reaches two PDMs around $\theta \sim$75$^\circ$. 
 This is the reason the trigger architecture is based on the
 PDM scale. The typical FoV of a PDM for $H_0 = 400$~km is about 
 30~km on a side ($\sim$1000~km$^2$). This means that the entire FS can be 
considered 
 as the sum of 137 quasi-independent sub-detectors corresponding to PDMs.
This is important for evaluating the effects of  
clouds and city lights. It should be mentioned here that when a 
trigger is issued on a PDM, the data of the neighboring PDMs are also 
retrieved.
Another important consideration
is that more inclined EASs will give
higher signals, either at EAS 
 maximum 
or as total integrated light. 
This can be used to 
 extend the energy range of measurement to lower energies by  simple geometrical cuts. Moreover, inclined showers will allow almost fully calorimetric 
measurement of the EAS because the entire profile will be visible. This is 
generally not the case of ground-based
detectors, which typically view up to 60$^\circ$, and for which the EAS is 
truncated at ground in many cases.

In conclusion, the three main players which define the trigger efficiency
for a specific night-glow background level and atmospheric conditions 
are the optics response
(most relevant), the zenith angle of the EAS (a factor of
2--3 in the collected light between quite inclined and vertical showers), 
and the distance effect ($\sim$25\% difference in the 
total number of photons reaching the pupil from the
same EAS located at the center or at the edge of the FoV).
Each of these three effects
can easily be identified by means of simple 
geometrical cuts on the zenith angle of the EAS and/or on its core location 
and accounted for given the intrinsic
characteristics of the detector. As an example, this means that the evolution 
of the exposure without geometrical cuts 
as a function of energy around the threshold 
can be verified by applying straightforward cuts on selected sub-samples of 
data where the aperture is known to be flat. This guarantees the quality of
the data even if the aperture without geometrical cuts  
has not reached the plateau yet.
 
 To include clouds in ESAF, a uniform and homogeneous layer is assumed. 
 Physical parameters considered for the cloud are obtained from a 
 `test cloud' layer defined
 by three input parameters: a) optical depth $\tau_{\rm C}$, b) 
 altitude $H_{\rm C}$, that yields a transmittance, 
 $\exp(-\tau_{\rm C})$ and c) physical thickness.
 
 \begin{figure}
   \vspace{5mm}
   \begin{center}
   \includegraphics[width=76mm]{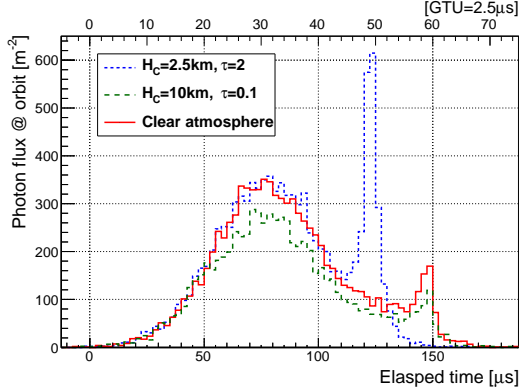}
   \end{center}
    \caption{Arrival time distribution of photons to
      the telescope per m$^2$ from an EAS of 
      $E = 10^{20}$~eV and $\theta = 60^\circ$
      with different cloud conditions.  
      Dashed and 
      dotted lines correspond to the cases of cirrus- and 
      stratus-like test clouds along with solid line 
      for the clear atmosphere case. } 
   \label{lupe-shower}
   \vspace{-0.5cm}
 \end{figure}

 In Figure~\ref{lupe-shower}, arrival time distribution of photons to the 
telescope of typical EAS
 events with zenith angle of $60^\circ$ are shown for cirrus- 
 ($H_{\rm C}$ = 10~km and $\tau_{\rm C} = 0.1$) and stratus- like 
 ($H_{\rm C}$ = 2.5~km and $\tau_{\rm C} = 2$) test clouds. For comparison, 
 the arrival time distribution of photons in clear atmosphere of the same
 EAS is shown.
 
 In case of a cirrus-like cloud at high altitudes, the signals from EAS 
 are attenuated according to the optical depth, while the EAS image and
 its time evolution
allow determination of the arrival direction.
The reflected signals of Cherenkov light from the landing surface are also 
observed.

 As mentioned in the previous section,  for stratus-like clouds with large 
 $\tau_{\rm C}$ at lower
 altitudes, most of the signal from EAS is observed without
 attenuation when  the cloud is well below the altitude of the shower 
 maximum. Such clouds also produce very intense
 reflected Cherenkov signals even larger than in the clear 
 atmosphere case. This may enhance the capability of triggering particular
 types of EAS such as low zenith angle events. Moreover, the
 reconstruction of the EAS geometry may benefit from such high reflectivity
 since the location of the impact on the cloud is more accurately 
 determined. On the other hand, a dedicated algorithm will be needed to 
 disentangle the contribution of Cherenkov light from fluorescence 
 light in estimating the energy of the event.

 \section{Geometrical aperture and cloud impact}
 \label{sec:aperture}
 
 \begin{figure}[!t]
   \vspace{5mm}
   \begin{center}
     \includegraphics[width=76mm]{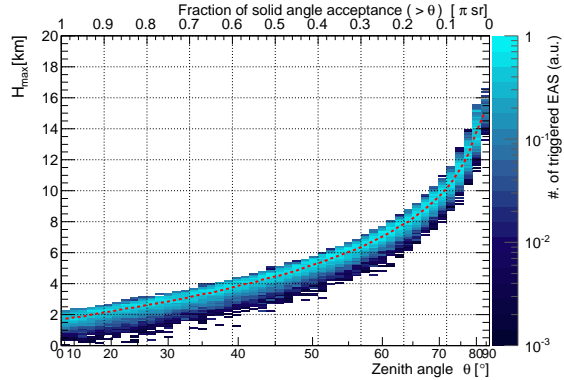}
   \end{center}
   \caption{Geometrical aperture as a function of energy.
     The filled circles and squares indicate geometrical apertures 
     for the entire observation area and $R<150$~km respectively,
     where $R$ indicates the distance of the impact location of the EAS
     from the center of FoV.
     The open circles and squares include a zenith angle cut 
     of $\theta >60^\circ$.}
   \label{fig2}
   \vspace{-0.5cm}
 \end{figure}
 \begin{figure*}[!t]
   \vspace{5mm}
   \begin{center}
     \includegraphics[width=76mm]{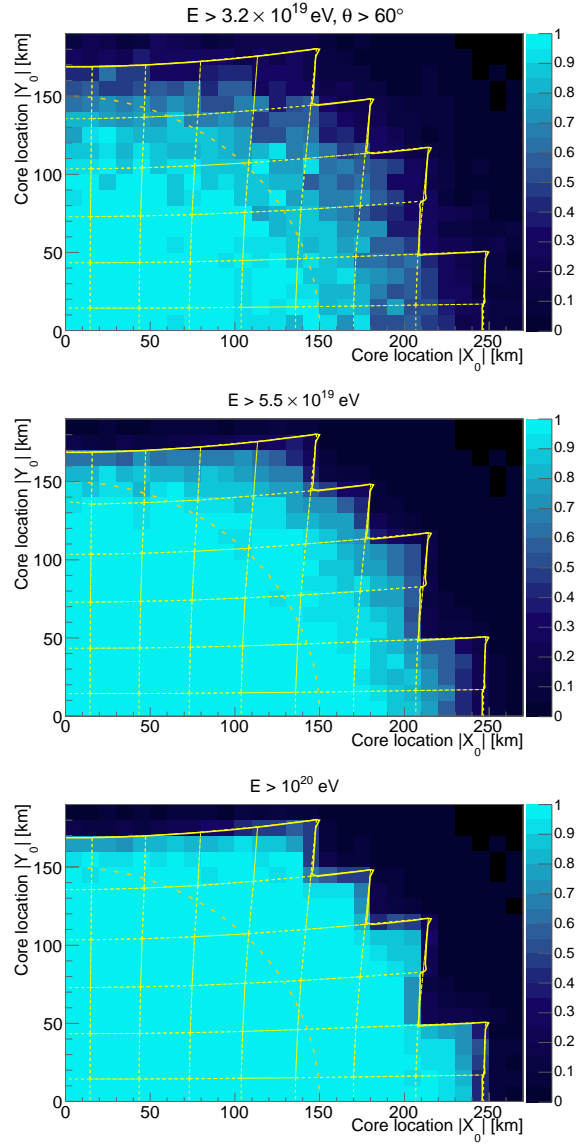}
   \end{center}
   \caption{Trigger probability as a function
     of core location with different geometrical and energy cuts: top)
     $E > 3.2\times 10^{19}$~eV with $\theta > 60^\circ$; middle)
     $E > 5.5\times 10^{19}$~eV and bottom) $E > 10^{20}$~eV. 
     The last
       two cases are  with no zenith angle cuts.}
     \label{trigger-maps}
 \end{figure*}

 To estimate the geometrical aperture, a large number of EASs are  
 simulated by uniformly injecting them over an extended area 
 $S_{\rm inject} \gg S_{\rm obs}$ in a clear atmosphere condition
 for nominal background level of $I_{\rm BG} = 500$ photons m$^{-2}$ 
~ns$^{-1}$~sr$^{-1}$. 
 
 For $N_{\rm trig}$, triggering samples among $N_{\rm inject}$, 
 simulated EAS events with an energy $E$, the corresponding geometrical 
 aperture $A(E)$ is defined by the following relation:

 \begin{equation}
   A(E) = {\displaystyle
     \frac{N_{\rm trig}} {N_{\rm inject}} \cdot S_{\rm inject}
     \cdot \Omega_0},
   \label{eqn:af}
 \end{equation}
 where $\Omega_0 = \pi$~sr is the solid angle acceptance for 
 0$^\circ$ $<$ $\theta$ $<$ 90$^\circ$. 
 As explained before, by applying simple cuts on the
 distance $R$ from the center of FoV of the impact location of the EAS, 
 and on the lower limit 
 $\theta_{\rm cut}$, the geometrical aperture $A_{\rm sub}$ is derived as 
 follows:
 \begin{equation}
   A_{\rm sub}(E) = 2\pi {\displaystyle \int_{S_{\rm sub}}
     \int^\pi 
     _{\theta_{\rm cut}}  \epsilon(E,\theta,\vec{r})
     \cdot\cos\theta \cdot \sin\theta d\theta  dS},
 \end{equation}
 where $dS$ is the area element in the selected subsection of the 
observation area $S_{\rm sub}$, and
 $\epsilon(E,\theta,\vec{r})$ is the probability of trigger at
 the impact location $\vec{r}$ with respect to the center of FoV.

 Figure~\ref{fig2} shows the geometrical aperture as a function
 of energy for $H_0= 400$~km along with the apertures for
 different geometrical cuts in $\theta$ and $R$. 
 Figure ~\ref{trigger-maps} shows the trigger efficiency as a function of 
core location for different cuts in $E$ and $\theta$.
 
 The geometrical aperture without geometrical cut 
 reaches the 
plateau\footnote{It is defined by the condition in which the geometrical 
   aperture is 
   $> 0.8 \cdot S \cdot \Omega$ for the area $S$ and 
   solid angle acceptance $\Omega$ defined by specific geometrical 
   cuts.}  above $\sim (6-7)\times  10^{19}$~eV. At the highest 
 energies, the geometrical aperture is close to saturation. The value is 
 mainly determined by $S_{\rm obs}$ for a given $H_0$ and, 
 therefore, higher altitudes result in the larger saturating apertures. Due to a minor contribution of EAS crossing the FoV, the geometrical aperture
grows slightly with energy.   

 By applying the cut $\theta > 60^\circ$, 
 which reduces the solid angle acceptance to $\pi/4$~sr, a constant aperture is 
 achieved above $\sim (4-5) \times  10^{19}$~eV. 
 In addition, a more stringent cut with 
 $R < 150$~km extends the constant aperture range down to $\sim 3\times 
 10^{19}$~eV. 
 The possibility to extend the plateau region at lower energies for a 
 subset of events will allow a cross-check of the flux measured by the full 
 sample of events in the specific range of energies where the 
 aperture of the instrument has not reached the plateau level yet. 
 Consequently, the overlapping energy range between JEM-EUSO and ground-based
 observatories will be enlarged.

\begin{figure}
  \vspace{5mm}
  \begin{center}
  \includegraphics[width=76mm]{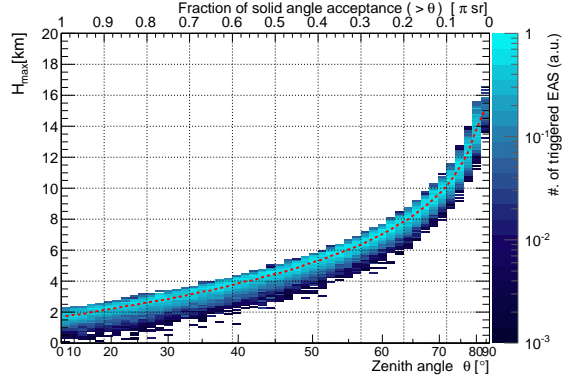}
  \end{center}
  \caption{Distribution of the altitude of EAS maximum $H_{\rm max}$ as a function 
    of zenith angle for triggered EAS events and an assumed differential flux 
    of $dN/dE \propto E^{-3}$. 
    The horizontal axis on the 
    top shows the solid angle acceptance above the given zenith angle.
    This result is obtained by merging together all triggered events of all
    energies simulated in clear atmosphere conditions.}
  \label{fig:EASmax-cloudtop}
\vspace{-0.5cm}
\end{figure}

 As mentioned in Section~\ref{sec:clouds}, the altitude of the EAS 
 maximum $H_{\rm max}$ compared to 
 that of the cloud-top 
 is an important parameter to decide if the EAS properties
 are reconstructed in a sufficiently precise way in presence of clouds. 
 Figure~\ref{fig:EASmax-cloudtop} represents $H_{\rm max}$ of triggered EASs 
 in clear atmosphere as a function of zenith angle. $H_{\rm max}$ is 
 strongly dependent on the zenith angle of the EAS. For proton EASs, 
 $\sim 80\%$ of the events have their maximum at altitudes higher than 3.2~km, 
 which is the value typically used in literature for the cloud-top altitude
 of low level clouds.  It should be mentioned that the 
 elongation rate of 
$X_{\rm max}$ (depth of shower maximum) , i.e. 
$\partial X_{\rm max}/\partial \log E$ is
 $\sim 80$ g cm$^{-2}$
per energy decade leading to only a limited increase 
$\partial H_{\rm max}/\partial \log E \sim 1$~km 
($\sim 0.3$~km) per decade for $\theta = 0^\circ$ ($60^\circ$). For 
heavier particles, $H_{\rm max}$ is slightly higher, and, 
therefore, less affected by the presence of clouds.

In order to evaluate the effect of clouds on the trigger
efficiency more precisely, 
EAS simulations for different cloudy cases are performed. Four 
cloud-top altitudes 
$H_{\rm C} = 2.5, 5, 7.5$ and 10~km are considered, as well as 
four optical depths 
of the test cloud $\tau_{\rm C}$= 0.05, 0.5, 1.5 and 5.  To 
quantify the effect of clouds, the ratio between the 
trigger aperture in a given cloudy condition to 
that of a clear atmosphere case, 
 $\zeta_{\rm C}(E; H_{\rm C},\tau_{\rm C})$, is calculated for each 
specific cloud
 condition $(H_{\rm C},\tau_{\rm C})$ as a function of $E$: 
\begin{equation}
  \zeta_{\rm C}(E; H_{\rm C},\tau_{\rm C}) = {\displaystyle 
    \frac{A(E; H_{\rm C},\tau_{\rm C})}{A(E;\rm clear)}}. 
 \end{equation}
\begin{table}
  \caption{Average $\zeta_{\rm C}$ values for 
    different types of clouds
    and EASs simulated with energies above $6.3\times  10^{19}$~eV
    and an assumed differential flux of $dN/dE \propto E^{-3}$.}
  \label{table_single}
  \begin{center}
    \begin{tabular}{l|cccc}
      \hline
      Cloud-top & \multicolumn{4}{c}{Optical depth $\tau_{\rm C}$}\\
      \cline{2-5}
      Altitude &0.05 & 0.5 & 1.5 & 5 \\
      \hline
      $H_{\rm C}=10$~km  & 90\% & 70\% & 26\% & 18\%\\
      $H_{\rm C}=7.5$~km & 89\% & 74\% & 43\% & 37\%\\
      $H_{\rm C}=5$~km   & 89\% & 82\% & 69\% & 66\%\\
      $H_{\rm C}=2.5$~km & 90\% & 88\% & 89\% & 88\%\\
      \hline
    \end{tabular}
  \end{center}
\end{table}
\begin{figure}
  \vspace{5mm}
\begin{center}
  \includegraphics[width=76mm]{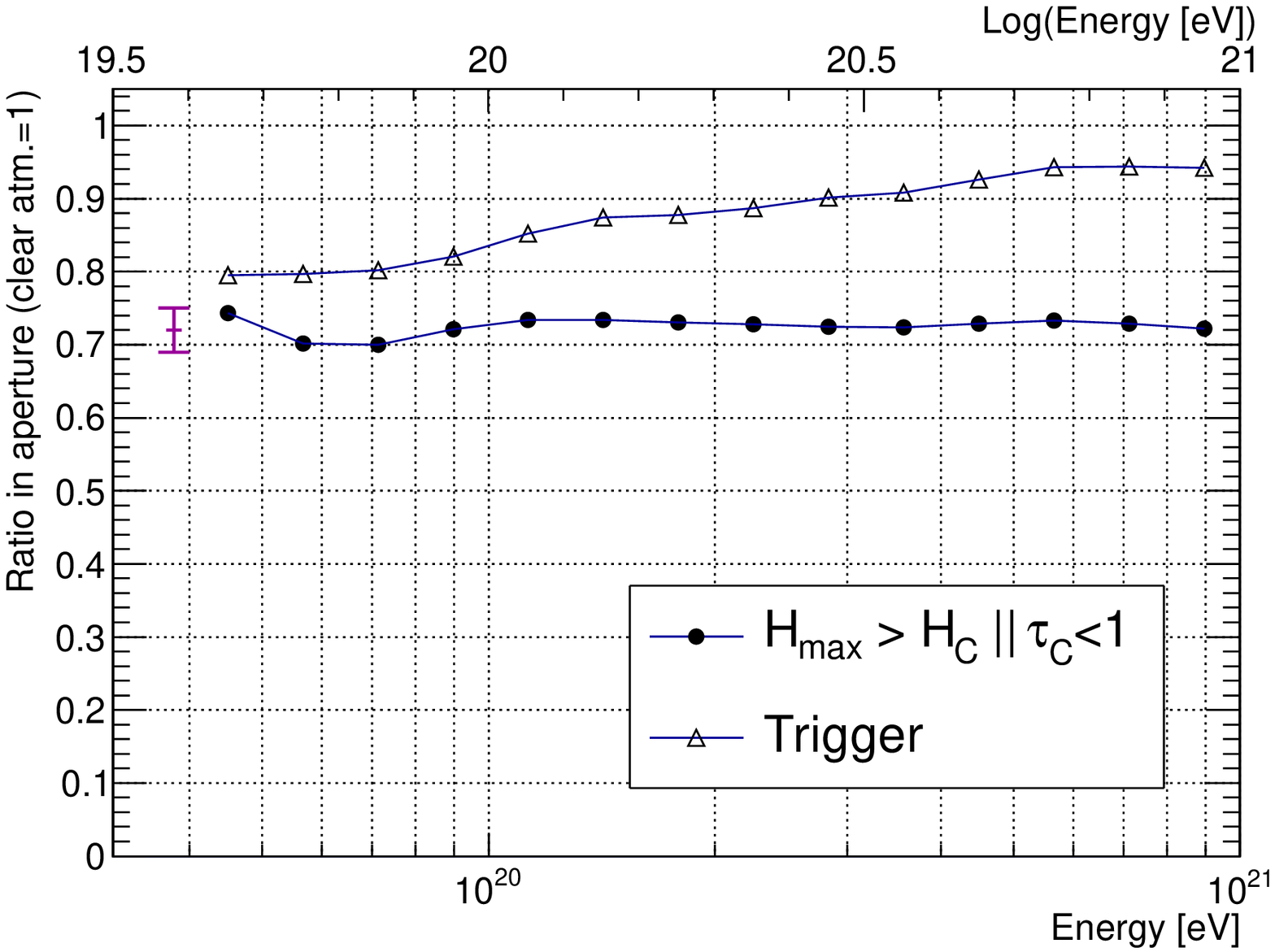}
  \end{center}
  \caption{Relation between cloud efficiency and energy. The triangles show, 
    as a function of energy, the results of the convolution of
    $\zeta_{\rm C}(E; H_{\rm C},\tau_{\rm C})$ of Table 
    \ref{table_single} with the matrix of cloud occurrence of Table~\ref{tab2}.
    Filled circles represent the case where we require
    $H_{\rm max} > H_{\rm C}$ 
    for optically thick clouds. The cloud efficiency 
    $\kappa_{\rm C}$ is defined as the average of the values expressed
    by the filled circles. The error bar on the left
shows the uncertainty on the points ($\sim$3\%). }
  \label{simp_fig2}
\vspace{-0.5cm}
\end{figure}
In Table~\ref{table_single}, the average values 
$\zeta_{\rm C}(E; H_{\rm C},\tau_{\rm C})$
are summarized for the different test clouds and showers simulated with 
energies
above $6.3\times 10^{19}$~eV and an assumed differential flux of 
$dN/dE \propto E^{-3}$.

In case of optically thick clouds with $\tau_{\rm C} \geq 1$,
the trigger efficiency depends on
$H_{\rm C}$. High altitude clouds in particular absorb EAS signals emitted 
beneath the cloud and result in a significant lowering of the trigger efficiency.
At middle altitudes, $H_C$ $\sim 5$~km, clouds only influence
EASs of small zenith angles, which develop at lower altitudes.

In the presence of high clouds with $\tau_{\rm C} < 1$, the signal
from an EAS below cloud level is only attenuated by a factor of
$\exp(-\tau_{\rm C})$ and the effect on the trigger efficiency is limited.
If $H_{\rm C}$ is well below the altitudes where EAS develop, the clouds
do not attenuate the signals. As a result of the different cuts, 
$\zeta_{\rm C}$ slowly increases with energy.

The results in Table~\ref{table_single} are then weighted with the 
relative cloud occurrence of Table~\ref{tab2} to estimate the average
effect as a function of energy. 
The results are shown in Figure~\ref{simp_fig2}.
For cloudy cases, the average of all triggered 
events is shown by triangles and it tends to increase with energy. 
The results obtained by applying the 
selection $H_{\rm max} > H_{\rm C}$ for $\tau_{\rm C} > 1$ cases 
is indicated by filled circles.
This cut requires that for optically thick clouds,
the EAS maximum is located above the cloud-top. This
ensures that fitting the EAS profile will not introduce significant distortion
of the reconstructed EAS profile.
With the above 
cut, the fraction of selected events over the ones triggering in clear-sky
conditions (the reference case) is almost constant at higher energies. 
This is because a certain
fraction of clouds with $\tau_{\rm C}>1$ exist at higher altitudes.
From Table~\ref{tab2}, for example, clouds with $H_{\rm C}>6.5$~km account
for $\sim$20\% of cloud coverage. Therefore, a part of EAS develops below 
such clouds. 
As this value is nearly constant as a function of energy, we define it as
the `cloud efficiency' $\kappa_{\rm C}$ and it accounts for
$\sim 72\%$ of the trigger EASs above $\sim 3 \times  10^{19}$~eV. 
This is due to the fact that the average $H_{\rm max}$ dependence is 
dominated by the zenith angle. The energy plays a smaller role.

The value $\kappa_{\rm C} \sim 72\%$ is an important factor for estimating 
the effective exposure of the mission. 
Currently, a detailed
study on the reconstruction of the events passing such a trigger selection
as well as those occurring in a clear atmosphere is in progress. It should be 
emphasized that
the main telescope of JEM-EUSO will be operated along with AM system
\cite{Andrii,Jose}.
  
\section{Exposure}
\label{exp-exp}
From the above results, the exposure per year of operation for events that
trigger JEM-EUSO, defined as the `annual exposure' 
is evaluated as a function of energy:
\begin{equation}
\mbox{(Annual exposure)}\equiv A(E) \cdot \kappa_{\rm C} \cdot \eta_0 
\cdot (1 - f_{\rm loc})\cdot \mbox{(1 [yr])}.
\quad 
\end{equation}
In this estimation, we use $\kappa_{\rm C} = 72\%, 
\eta_0 = 20\%,$ and $f_{\rm loc} = 10\%$, respectively.
The operational inefficiencies related to ISS (rockets docking on ISS,
lid operation, detector maintenance or aging, etc.) as well as
quality cuts on reconstruction are not taken into account yet, and will be
addressed in future. Therefore, the present results constitute
an upper limit on the effective exposure of the instrument for the
assumed conditions.
\begin{figure}
  \vspace{5mm}
  \centering
  \includegraphics[width=76mm]{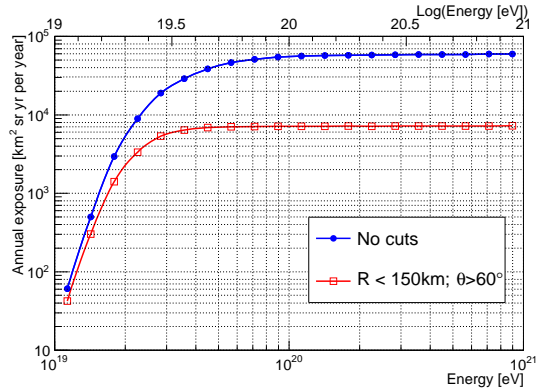}
  \caption{Annual exposure as a function of energy obtained for
the two extreme conditions shown in Figure~\ref{fig2}: a) entire
observation area (filled circles); b) cut on distance $R<150$~km
and on zenith angle $\theta >60^\circ$ (open squares). 
}
  \label{fig224}
\vspace{-0.5cm}
\end{figure}

In Figure~\ref{fig224}, the annual exposure as a function of energy is shown.
The filled circles indicate the geometrical aperture
for the entire observation area.
The open squares include a zenith angle cut
of $\theta >60^\circ$ and a cut on distance of $R<150$~km. These two exposures
correspond to the highest and lowest aperture curves of Figure~\ref{fig2}.

The JEM-EUSO annual exposure for the full sample of data is
expected to be $\sim$ 9 times larger than that of the Pierre Auger
Observatory with the corresponding annual exposure of about
7000~km$^2$~sr yr, at energies around $10^{20}$~eV.
Because of the steeply rising aperture at lower energies, the subsets of
data with reduced and flat exposure will be used to cross-check  
with measurements by other ground-based experiments
down to $\sim (2-3)\times 10^{19}$~eV.
It is important to underline that the cuts shown in Figure~\ref{fig224} 
are extreme in order to obtain at $\sim 3\times 10^{19}$~eV an annual 
exposure comparable to that of Auger, which means 
acquiring a statistically 
similar data sample. As shown in Figure~\ref{fig2} for the
apertures, less stringent geometrical cuts will lead to different 
exposure curves located in between
the two lines shown in Figure~\ref{fig224} with a flat plateau starting 
gradually at higher energies.  
In this way, it will be possible to have a comparison of UHECR fluxes for 
one entire decade
in energy using the data acquired 
without geometrical cuts.


It is worthwhile remembering here that the aperture and 
exposure have been derived with specific assumptions on the detector 
properties, background level, shower development in atmosphere, etc. All
the systematic uncertainties that would increase or 
decrease the collected 
light at telescope level, either for the EAS or for the background, would 
be responsible to shift the energy scale of the aperture and exposure 
curves by the square root of the systematic uncertainty. 
On the other hand, the
scaling factor would be linear in case it involves only the EAS 
propagation (i.e. hadronic interaction model, fluorescence yield).

In the previous analysis, a constant background level of 
$\langle I_{\rm BG}\rangle = 500$~photons m$^{-2}$~ns$^{-1}$~sr$^{-1}$ 
was assumed. 
However, the background is variable with time.
To take into account the effective background variation, the 
exposure 
over the time when $I_{\rm BG} < I_{\rm BG}^{\rm thr}$
given as a function of $E$ is approximated by the 
following relation:
\begin{equation}
  \mbox{(Overall exposure)}
  \propto 
  {\displaystyle \int^{I_{\rm BG}^{\rm thr}}_0 A
    \left(\sqrt{\frac{\langle I_{\rm BG}\rangle}{I_{\rm BG}}} \cdot E \right)
  \cdot p(I_{\rm BG}) dI_{\rm BG}}.
  \label{eqn:conv}
\end{equation}
\begin{figure}
  \centering
  \includegraphics[width=76mm]{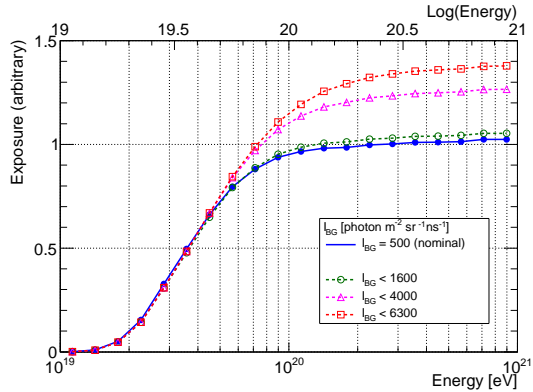}
  \caption{Exposure in arbitrary unit for different values of
    maximum acceptable background level. 
}
  \label{bg-THRESHOLD}
  \vspace{-0.5cm}
\end{figure}

Figure~\ref{bg-THRESHOLD} shows the exposure as a function of $E$ 
for various maximum allowed background levels obtained by 
convolving the trigger probability at a specific fixed background 
level with the fraction of time during which such background level occurs 
according to the estimation by Equations~(\ref{eqn:obsduty}) and 
(\ref{eqn:moonshine}). As described in Section~\ref{trigger}, 
the trigger system is capable of dynamically adjusting the thresholds to 
cope with variable background intensity.
The trigger efficiency curve scales in energy 
approximately in a proportional way to 
$\sqrt{I_{\rm BG}}$ because it depends on the Poissonian
fluctuations of the average background level.

The exposure obtained with a fixed $I_{\rm BG}$ of 500 
photons m$^{-2}$~ns$^{-1}$~sr$^{-1}$ 
is essentially equivalent to the one obtained from Equation~(\ref{eqn:conv})
when the integration of $I_{\rm BG}$ is extended up to 1600 photons 
m$^{-2}$~ns$^{-1}$~sr$^{-1}$. 
It is possible to observe that 
at higher energies there is still some margin
of gain if a higher level of background is accepted (see i.e. the curve at 
less than 6300 photons m$^{-2}$~ns$^{-1}$~sr$^{-1}$). This is particularly 
useful to explore the extreme energy ranges where 
the flux is rapidly decreasing with energy. In any case, all
the conclusions obtained in this paper are derived assuming 
only the standard condition of 500 photons m$^{-2}$~ns$^{-1}$~sr$^{-1}$ 
constant background level. 

Unlike ground-based observatories, the global ISS orbit
and better sensitivities for EAS with large zenith angles allows observation
of the entire Celestial Sphere. The exposure distribution  is
practically flat in right ascension. Apart from possible local
or seasonal deviation from the global average of cloud coverage
and of background level, the relationship between the expected overall 
exposure and declination can be analytically calculated as a function of
only $\theta_{\rm cut}$, knowing the observable nighttime at a
given latitude.

\begin{figure}
  \begin{center}
    \includegraphics[width=76mm]{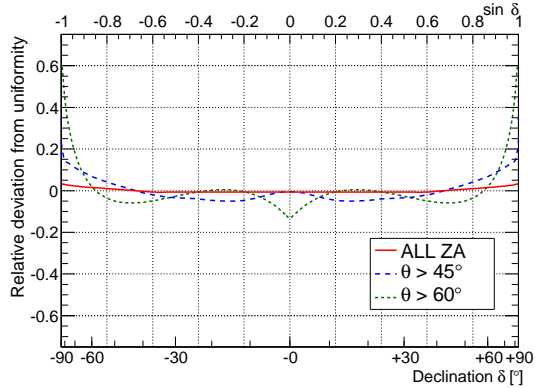}
  \end{center}
  \caption{Expected distribution of observed exposure as a function
    of declination with different zenith angle cuts for  all
    zenith angles (solid line), 
    $\theta > 45^\circ$ (dashed line) and $\theta > 60^\circ$ (dotted line).
    The vertical axis 
    indicates the deviation from the uniform distribution.}
  \label{fig3}.
\vspace{-0.5cm}
\end{figure}

Figure~\ref{fig3} shows the expected distribution of observed exposure 
as a function of declination on the celestial sphere 
with different zenith angle cuts for all zenith angles (solid line), 
$\theta > 45^\circ$ (dashed line) and $\theta > 60^\circ$ (dotted line ). 
The vertical axis indicates the deviation from the uniform distribution.

For the case of $\theta > 60^\circ$, minor
excesses and deficits arise in very limited regions near
Celestial Poles and Equator, respectively.  This is because the
ISS has a slightly longer residence time at
high latitudes. JEM-EUSO can achieve in general a nearly constant
exposure for the full range of zenith angles for which the arrival
direction analysis will be performed. This is one of the advantageous features
of space-based observation, as only small correction factors are needed.

\section{Discussion}
\label{discussion}

The convolution of observational duty cycle $\eta_0$ $\sim$20\%, cloud 
efficiency
$\kappa_c$ $\sim$72\% and effect of local light 
$f_{\rm loc}$ $\sim$10\%
 gives an overall conversion factor from geometric aperture to exposure 
of about $\sim$13\% with slight variations depending on the cuts applied 
to the different terms. 
This is due to the particular features of observation
from space: a) the possibility of operating in the presence of 
some diffuse moonlight up to 68\% of the time, and b) the 
possibility of measuring in a substantial fraction of the solid angle 
acceptance
even in presence of low- or middle- altitude clouds. 
As an example, the distribution of cloud
occurrence around the location of Auger has been compared to the ISS case
by using TOVS data though with statistical errors $\sim 6\%$. 
Results show that, despite a 
significantly higher fraction of clear atmosphere at Auger site 
($\sim 42\%$) compared to the world average (day and night) spanned by 
the ISS ($\sim 28\%$), when the fraction of time in which the
cloud-top is located below 3.2~km is added to the clear atmosphere, both 
Auger and ISS orbit give
similar results ($\sim 60\%$). For space-based observation 
this is the factor that should be compared with the clear atmosphere 
fraction of a ground-based observatory. 

Our results are in 
agreement with those obtained by
\cite{ESAFpaper} for the observational duty cycle. 
Our result is slightly higher than what has been reported in \cite{Mitchell}
for the OWL observatory mainly due to the fact that OWL will fly at higher 
altitudes ($\sim$ 1000~km),
reducing the duty factor for full darkness by about 15\%--20\% due 
to the shorter nighttime.
Moreover, in our estimation we have taken into account the possibility of
accepting some moonlight as long as the total diffuse light is less than
1500 photons m$^{-2}$~ns$^{-1}$~sr$^{-1}$.

Concerning clouds, our efficiency is a factor of  $\sim 1.5$ times larger 
than \cite{ESAFpaper} essentially due to the fact that the cut on 
$H_{\rm max}$ is applied only 
to optically thick clouds. 
This is motivated by the fact that thin clouds might distort the shower
profile but not affect the arrival direction of the primary
particle and therefore may be used to perform anisotropy 
analyses.

Even though it is not considered in the present analysis,
it is interesting to observe that optically thick clouds 
($\tau_{\rm C} > 2$), located
at low altitudes, may play another positive role, which is blocking the 
anthropogenic light and, therefore, allowing EAS measurement also 
in regions typically polluted by city light.

Our results do not support the conclusions in \cite{Sokolsky}, 
where it was claimed that the sensitivity of space-based 
fluorescence detectors is of unacceptably small level.
In that approach, only cloud-free
scenes are considered and we know from Section~\ref{sec:clouds}  
that they account for only 1/3 of the time. Moreover,
very strict conditions were applied on the extension of the cloud-free 
area. As an example, no
correlation on the altitude of the possibly cloudy pixel and location of 
the track was applied. This tends to reject very
inclined events with much longer paths in atmosphere even though
all the detectable part of the track is located above clouds.
It is clear that these constraints severely reduce the overall
efficiency. The authors commented that 
less rigorous constraints could increase
the cloud-free efficiency by even a factor of 3.
Finally, as previously mentioned, simulations have proved 
\cite{Takahashi,AbuZayyad}
the feasibility of reconstructing EAS with reasonable uncertainty in presence
of clouds. It is important to
stress that the AM system will have an important role in
monitoring the atmospheric conditions in which EAS develop, together
with information from satellites, ground-based observations
and meteorological models.
We are, therefore, confident that it will be possible to characterize
the atmospheric conditions in which each EASs 
has been detected, and, account for these conditions in the data analyses.
Finally, it has to be mentioned that most of the above problems are 
related to the estimation of the energy and $X_{\rm max}$ of the EAS.
The estimation of the exposure is more easily assessed
and does not depend strongly on the level of accuracy of the EAS 
reconstructed parameters.

The present results indicate that JEM-EUSO has the potential to 
reach an annual exposure of nearly one order of magnitude higher than 
Auger at energies around 10$^{20}$~eV. 
However, a final assessment of the
annual exposure requires an evaluation of event selection efficiency  
to ensure the quality of the reconstructed
EASs in terms of energy, angular resolution, and $X_{\rm max}$.
This will be described in detail in a forthcoming paper. 
Preliminary results (see \cite{Bertaina}), assuming clear
atmosphere indicate that this condition
is satisfied by most of the events.

It is possible that the full data sample will be 
subdivided in sub-samples of data peculiar for each analysis. 
As an example, those events reconstructed in thin cloud conditions are
still most probably usable for the anisotropy analyses, granted that
a lower limit on the energy of the event is assigned with high
confidence. 
Moreover, the possibility of defining a sample of events detected in
`golden conditions' such 
as inclined EASs in clear-atmosphere will guarantee a 
cross-check of the reliability of wider samples of events.

It will be possible to significantly increase the exposure by
tilting the telescope.
In the tilt mode, the observation area is scaled by $\sim (\cos\xi)^{-3}$
as a function of titling angle $\xi$ of the optical axis from the nadir. 
This will increase the sample at the highest energies and help to compensate
the reduction of the observation area in case of periods of lower orbiting 
altitude. As an example, in case of $H_0 \sim 350$~km, tilting 
the instrument by $\xi \sim25^\circ$ would give an observation area 
similar to the case of $H_0 \sim 400$~km in nadir mode. At the 
same time, the 
observation in nadir mode would extend by $\sim$30\% the lowest 
energies where the measurement would be feasible.
The analyses in such `quasi-nadir mode' in which the optical axis is 
tilted by $\sim 0^\circ-25^\circ$, can be easily assimilated to the  
nadir one. In case of even larger tilting angles ($\xi \ge 25 ^\circ$), 
a dedicated study is necessary to evaluate the performance. This
will be addressed in future.

We wish also to point out that JEM-EUSO has considerably improved with 
respect to the original Extreme Universe Space Observatory \cite{EUSO} 
that successfully completed Phase-A study within ESA. 
The main improvements can be ascribed to the baseline optics 
of the JEM-EUSO telescope \cite{Alex} 
(with 
$\sim$1.5 better focusing capability), 
to the FS detector 
\cite{Kawasaki} ($\sim$1.6 higher detection 
efficiency), to the better geometrical layout of PDMs on the FS that 
maximizes the filling  factor \cite{MRicci}, and to the improved 
performance of the electronics \cite{Il,Joerg}, which allow more 
complex trigger algorithms \cite{Bertaina2}.

\section{Summary}
\label{summary}

The most important factors which determine
the annual exposure of JEM-EUSO mission have
been reviewed. The analytical calculations indicate that the
operational duty cycle of JEM-EUSO, or the fraction of time
in which the EAS measurement is not hampered by the brightness
of the atmosphere, is of the order of $\eta_0$ $\sim$20\%. 
The local light such as city light, atmospheric flashes and auroras will 
reduce the effective instantaneous observational area to $1-f_{\rm loc} 
\sim 90\%$ of the geometrical area. The role of clouds has been thoroughly 
investigated and the cloud efficiency, defined as the ratio of the 
effective average aperture to the geometrical aperture, is found to be 
$\kappa_c$ $\sim$72\%. All the above factors give an overall conversion 
factor from geometric aperture to exposure of about $\sim$13\%.
Simulations show that JEM-EUSO can reach almost full efficiency at 
energies around $3\times 10^{19}$~eV for a restricted subset of events, 
and for the full aperture at energies $E\gtrsim (6-7)\times 10^{19}$~eV. 
The expected annual exposure of JEM-EUSO around 10$^{20}$~eV is equivalent 
to about 9 years exposure of Auger. This value has to be presented as the 
potential of JEM-EUSO in nadir mode. A study of the selection efficiency 
of events due to quality cuts on reconstruction in clear and cloudy 
conditions will be performed in future to refine these results. 

\section*{Acknowledgments}
This work was partially supported by Basic Science Interdisciplinary 
Research Projects of RIKEN and JSPS KAKENHI Grant (22340063, 23340081, and 
24244042), by the Italian Ministry of Foreign Affairs, General Direction 
for the Cultural Promotion and Cooperation, by the 'Helmholtz Alliance 
for Astroparticle Physics HAP' funded by the Initiative and Networking Fund 
of the Helmholtz Association, Germany, and by Slovak Academy  
of Sciences MVTS JEM-EUSO as well as VEGA grant agency project 2/0081/10.
The Spanish Consortium involved in the JEM-EUSO Space
Mission is funded by MICINN under projects AYA2009-06037-E/ESP, 
AYA-ESP 2010-19082, AYA2011-29489-C03-01, AYA2012-39115-C03-01, 
CSD2009-00064 (Consolider MULTIDARK) and by Comunidad de Madrid (CAM) 
under project S2009/ESP-1496.

The many constructive comments and suggestions
of two anonymous referees are acknowledged.

\end{document}